\documentclass[1p, preprint]{elsarticle}
\usepackage[utf8x]{inputenc}
\usepackage[english]{babel}
\usepackage{amssymb}
\usepackage{amsmath}
\usepackage{multirow}
\usepackage{graphicx}
\usepackage{leftidx}
\usepackage{url}
\usepackage{amsmath}
\DeclareMathOperator{\Tr}{Tr}

\begin{document}

\title{Recent advances in percolation theory and its applications}

\author[]{Abbas Ali Saberi}
\ead{ab.saberi@ut.ac.ir, $\&$ ab.saberi@gmail.com}

\address[]{Department of Physics, University of Tehran, P.O. Box 
14395-547,Tehran, Iran}
\address[]{School of Particles and Accelerators, Institute for Research in Fundamental Sciences (IPM) P.O. Box 19395-5531, Tehran, Iran}

\begin{abstract}
Percolation is the simplest fundamental model in statistical mechanics that 
exhibits phase transitions signaled by the emergence of a giant connected 
component. Despite its very simple rules, percolation theory has successfully 
been applied to describe a large variety of natural, technological and social 
systems. Percolation models serve as important universality 
classes in critical phenomena characterized by a set of critical exponents 
which correspond to a rich fractal and scaling structure of their 
geometric features. We will first outline the basic features of the ordinary 
model.

Over the years a variety of percolation models has been introduced some of 
which with completely different scaling and universal properties from the 
original model with either continuous or discontinuous transitions depending on 
the control parameter, dimensionality and the type of the underlying rules and 
networks. We will try to take a glimpse at a number of selective variations 
including Achlioptas process, half-restricted process and spanning 
cluster-avoiding process as examples of the so-called explosive percolation. We 
will also introduce non-self-averaging percolation and discuss correlated 
percolation and bootstrap percolation with special emphasis on their recent 
progress. Directed percolation process will be also discussed as a prototype of 
systems displaying a nonequilibrium phase transition into an absorbing state.

In the past decade, after the invention of stochastic L\"{o}wner evolution 
(SLE) by Oded Schramm, two-dimensional (2D) percolation has become a central 
problem in probability theory leading to the two recent Fields medals. After a 
short review on SLE, we will provide an overview on existence of the scaling 
limit and conformal invariance of the critical percolation. We will also 
establish a connection with the magnetic models based on the percolation 
properties of the Fortuin-Kasteleyn and geometric spin clusters. As an 
application we will discuss how percolation theory leads to the reduction of 
the 3D criticality in a 3D Ising model to a 2D critical behavior.

Another recent application is to apply percolation theory to study the 
properties of natural and artificial landscapes. We will review the statistical 
properties of the coastlines and watersheds and their relations with 
percolation. Their fractal structure and compatibility with the theory of SLE 
will also be discussed. The present mean sea level on Earth will be shown to 
coincide with the critical threshold in a percolation description of the global 
topography. 
\end{abstract}

\begin{keyword}
  percolation \sep explosive percolation \sep SLE \sep Ising model \sep Earth topography
\end{keyword}

\maketitle
\pagenumbering{roman}
\tableofcontents

\pagenumbering{arabic}

\section{Introduction}
\label{intro}

\begin{figure}[t]
\centerline{
\includegraphics[width=0.6\textwidth]{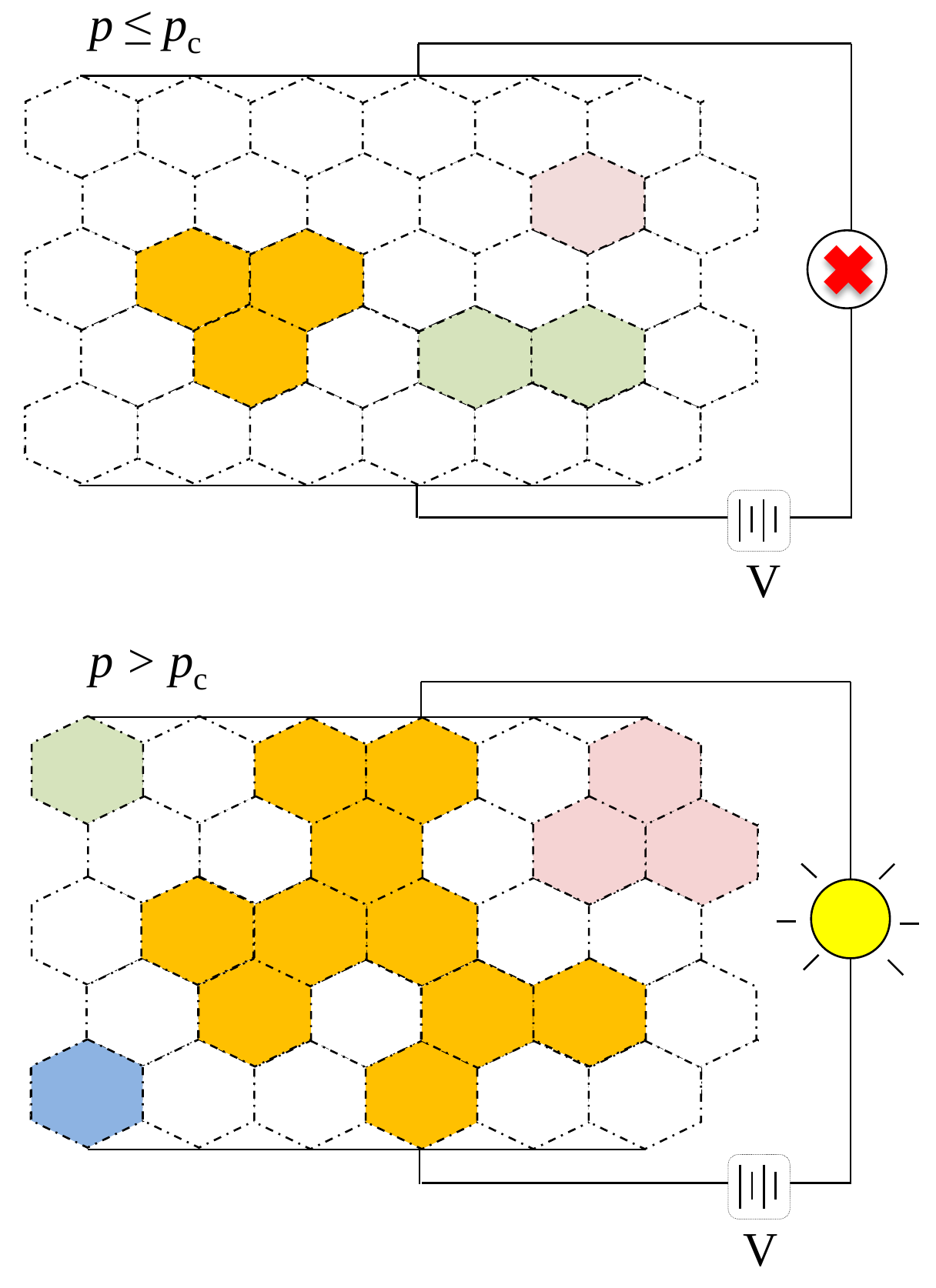}
}
 \caption{Schematic illustration of a simple circuit including a bulb and a voltage source connected to an insulator honeycomb lattice which can be furnished by a fraction $p$ of metallic plaquettes. Once $p$ reaches a critical threshold $p_c$, a spanning metallic cluster emerges which connects the two sides of the lattice and the bulb suddenly lights up.}
\label{fig1}
\end{figure}

Consider a simple electric circuit consisting of a voltage source and a bulb 
connected to an insulating hexagonal honeycomb lattice of size $N$. Now imagine 
we start occupying the lattice in a random way by $n$ number of metallic 
hexagonal plaquettes. For small values of fraction of occupation $p=n/N$, lower 
than a critical threshold $p_c$, there appear some small metallic clusters but 
the two opposite sides of the lattice still remain disconnected leading to the 
lack of charge flow in the circuit and thus the bulb remains off---see 
Fig.~\ref{fig1}.  A metallic cluster is defined as a set of occupied sites that
can be traversed by jumping from neighbor to occupied neighbor. In 
Fig.~\ref{fig1} the different colors simply label the different isolated 
clusters and have no other significance. By increasing the number of the 
randomly distributed metallic plaquettes (and increasing the occupation 
probability $p$ indeed), the average size of the metallic clusters increases. 
Once it exceeds a certain threshold value $p_c$, a spanning cluster emerges 
that closes the circuit and the bulb suddenly lights up. This transition from 
insulator to the metallic phase in two dimensions (2D) exemplifies one of the 
simplest and fundamental classes of phase transitions in statistical physics 
called "percolation," which can generally be defined in any dimension $d$. 

The idea of percolation model was first effectively considered by Flory in 
1940s \cite{flory1941molecular1, flory1941molecular2, flory1941molecular3}. 
However, the study of the model as a mathematical theory, dates back to 1954 
\cite{CahnNature97}, when engineer
Simon Broadbent and mathematician John Hammersley, one concerned with the design of carbon filters for gas masks, put their heads together to deal with “The stepwise spreading of a fluid or individual particles through a medium
following a random path in which each link is either open or shut, according to a specified statistical proportion. The spreading process is therefore arrested at many sites...”\cite{Broadbent1957}.
A long path along which spreading is not interrupted constitutes an infinite (or spanning) cluster. 
Broadbent and Hammersley proposed the concept of a percolation threshold above which the links form an infinite cluster with high probability. 

Percolation theory was then popularized in the physics community and intensively studied by physicists \cite{essam1961some, fisher1961statistical, kirkpatrick1973percolation, stauffer1979scaling, essam1980percolation, Isichenko1993Percolation, sahimi1993percolation, stauffer1994percolation, sahimi1994percolation, Bunde1996percolation, King2002percolation, Stauffer2009Review}. It has been found to have a broad applications to diverse problems as understanding conducting materials \cite{vigolo2005experimenta, grimaldi2006tunneling}, the fractality of coastlines \cite{Sapoval2003, Saberi2013PRL}, networks \cite{Derenyi2005, Callaway2000, Kalisky2006}, turbulence \cite{Cardy2006Nature, Bernard2006Nature}, magnetic models \cite{Fortuin1969, Fortuin1972, Dotsenko95NPB, Dotsenko93PRL, SaberiEPL}, colloids \cite{Anekal2006, Gnan2014}, growth models \cite{SaberiAPL}, retention capacity and watersheds of landscapes \cite{Knecht2012, Baek2012PRE, SchrenkZiff}, the spin quantum Hall transition \cite{Gruzberg1999PRL} and SU(3) lattice gauge theory \cite{Endroedi2014}. From a mathematical
point of view percolation is also attractive because it exhibits relations between probabilistic and algebraic/topological properties of graphs. Although a significant amount of research has been done in the field there still exist many unsolved problems \cite{Ziff2014Recent}.

One of the active areas in percolation research which has still open problems is to find the percolation thresholds $p_c$, as a fundamental characteristic of
percolation theory, both exactly and by
simulation. The values of percolation thresholds are not universal and 
generally depend on the structure of the lattice and dimensionality, and are 
believed to achieve their mean-field values only in the limit of infinite 
dimension \cite{essam1961some}. Finding mathematical rigorous proofs for 
thresholds and bounds has also been of research interest in 
\cite{kesten1980critical, wierman1984bond, Grimmett1999percolation}. Exact 
thresholds
in 2D for the square, triangular, honeycomb
and related lattices 
were found using the star-triangle transformation \cite{Skeys1963}. It has been shown in \cite{Ziff2006JPA} 
that thresholds can be found for any lattice that can be represented as a self-dual 3-hypergraph (that is, decomposed into triangles that form a self-dual arrangement).
It is also shown in \cite{Grimmett2013PTRF} that
for any lattice which can geometrically be represented as an isoradial graph, 
it is possible to find the thresholds. This leads to a wide class of exact 
thresholds and provides a proof
\cite{Ziff2012JPA} of the conjecture \cite{Wu1979JPC} for the threshold of the
checkerboard lattice. However, the exact value of thresholds for many systems 
of long interest are still missing \cite{Ziff2014Recent}.

\begin{figure}[h]
\centerline{
\includegraphics[width=0.6\textwidth]{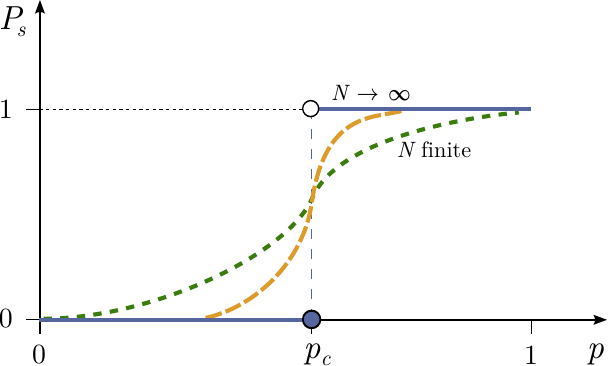}
}
 \caption{Schematic plot of the percolation probability as a function of the occupancy. As the system size $N$ goes to infinity, $P_s$ tends to the step function around the critical threshold $p_c$. For very special systems all graphs for finite $N$ cross at a single point, but in general, because of finite-size corrections, the crossing does not occur at a single point for finite systems \cite{Reynolds1980PRB, ZiffNewman2002}.}
\label{fig2}
\end{figure}

For our mentioned example at the beginning of the section, it is equivalent to a 
2D \textit{site percolation} problem on a triangular lattice whose sites are 
placed at the center of the plaquettes with six number of nearest neighbors. 
The value of the threshold for this case is exactly known to be $p_c=1/2$ in 
the infinite system size limit. In fact, if our system size was infinitely 
large or equivalently, we could have infinitely small plaquettes, then we would 
need $N\longrightarrow\infty$ number of plaquettes to totally furnish the 
lattice. In that case, the percolation probability defined as the probability 
to have an infinite cluster~\footnote{However, one should be careful with the 
definition of "infinite".}, would be a step function around $p_c$ which  takes 
the value 1 for $p>p_c$ and zero if $p\leq p_c$---see Fig.~\ref{fig2}. At 
$p=p_c$, it is believed that with probability one there is no infinite cluster, 
but there are typically some very large clusters nearby \cite{Aizenman1997}.\\
One may notice that in our example above, we could instead have occupied the edges of the sites (or the bonds) with some metallic rods rather than occupying the plaquettes themselves.
This would then change the problem to one that is known as \textit{bond percolation} on a 2D honeycomb lattice with three number of nearest neighbors. This approach changes the percolation
threshold to $1-2\sin(\pi/18)\approx 0.65271$ \cite{Skeys1963}, but does not affect the other fundamental properties, we shall return to this point later. \\The other point is that in reality the insulator lattice in our experiment is not infinite of course, and it has a finite size. The problem of how to deal with finite size lattices is known as \textit{finite size scaling}. It is
a useful introduction to the style of theoretical argument that is often used in
percolation theory \cite{King2002percolation}. In such situations if we repeat 
our experiment many times for a given occupancy $p$ with different realizations 
of randomness to estimate the percolation probability, we would find a smooth 
function as represented in Fig.~\ref{fig2}, rather than a step function around 
$p_c$. This means that in reality we can get connectivity even at very much 
less than the percolation threshold or not get it even at a much higher 
occupancy. As the size of the system
gets larger the scatter around the sharpness would reduce until we return to the plot for the infinite system size.

\begin{figure}[h]
\centerline{
\includegraphics[width=0.55\textwidth]{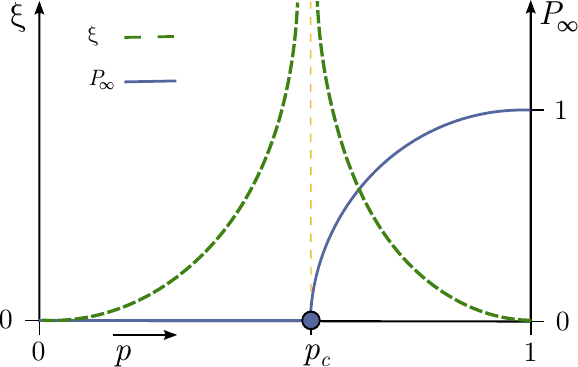}
}
 \caption{Schematic diagram of the  percolation strength $P_\infty(p)$ and the correlation length $\xi(p)$ as a function of the occupancy $p$ for a system of infinite size.}
\label{fig3}
\end{figure}

The sudden onset of a spanning cluster at a particular value of the occupation probability along with a number of characteristic features make the percolation transition a nontrivial critical behavior. This criticality belongs to a large family of critical phenomena with common remarkable features and forms an important universality class characterized by a number of scaling laws and critical exponents. The notion of universality means that the large scale behavior of critical systems can be
described by relatively simple mathematical relationships which are entirely
independent of the small scale construction. This property enables one to study 
and understand the behavior of a very wide range of systems without needing to 
know much about the details. For example, the universal properties of the 
percolation model are entirely independent of the type of lattice (e.g., 
hexagonal, triangular or square, etc.) or whether it is site or bond 
percolation; they only depend on the dimensionality of the system.\\ Another 
example of a prototype model in critical phenomena is the Ising model as a 
mathematical model of a magnet which undergoes a phase transition between a 
ferromagnetic ordered phase and a paramagnetic disordered phase at a particular 
temperature $T_c$, the Curie temperature, in more than one dimension. The net 
magnetization which is the first derivative of the free energy with respect to 
the applied magnetic field strength, is the order parameter which distinguishes 
between these two phases i.e., it takes the value $0$ in the paramagnetic phase 
and increases continuously from zero as the temperature is lowered below the 
Curie point. Such transitions in which the order parameter is continuous but 
second derivative of the free energy (like magnetic susceptibility and heat 
capacity in the Ising model) exhibits a discontinuity are called second-order 
or continuous phase transitions. First-order (discontinuous) transitions 
involve a discontinuous change in their order parameter. Both percolation and 
Ising models display a continuous phase transition and as will be made clear 
later in Sec.~\ref{chap4}, they are closely related to each other 
\cite{Fortuin1969, Fortuin1972}. Therefore we have to be able to define an 
order parameter for the percolation model as in the Ising model. It seems not 
to be so difficult: not all occupied sites are in the infinite (or spanning) 
cluster, thus if we look at the
probability $P_\infty(p)$ that an occupied site is in the infinite cluster for a given occupancy $p$, then this must be zero (since there is no spanning cluster) below the percolation threshold and increases continuously (but of course with singular derivatives) as one enters the supercritical (connected) phase \footnote{The system is called to be in the subcritical (resp.
supercritical) phase if $p < p_c$ (resp. $p > p_c$).}---see Fig.~\ref{fig3}. 
Continuity of the strength probability $P_\infty(p)$, as the main macroscopic 
observable in percolation, at $p_c$ is an
open mathematical problem in the general case, but it is known to
hold rigorously in 2D and $d\geq19$ \cite{Hara1994Meanfield} using lace expansion methods. The conjecture
that $P_\infty(p=p_c)=0$ for $3\leq d\leq18$ remains however one of the
open problems in the field \cite{Beffara2008Percolation}.
  
Percolation is a simple model of robustness and stability exhibiting a robust continuous transition in all dimensions. It accounts as
a fundamental step in dealing with more complex models and even dynamical processes occurring on
the networks. There have been introduced various modifications of the ordinary 
percolation to model different statistical phenomena or even at the level of 
mathematical curiosity in building up a modified version with a possibly 
discontinuous phase transition \cite{Chalupa1979, Alder1991, Bollobas1984, 
Dorogovtsev2006}. Explosive percolation \cite{Achlioptas2009Explosive}, for 
example, is one of the most challenging cases with a seemingly mild 
modification of standard percolation model which was first, surprisingly, 
claimed to exhibit a discontinuous phase transition, in contrast to the 
ordinary percolation. It then has been realized that the finite size effects 
make it difficult to numerically realize if the explosive percolation in random 
graphs is continuous or discontinuous \cite{Hermann2010Explosive, ziff2009PRL, 
Cho2009PRLPercolation, Friedman2009PRL, DSouza2010PRL, NaglerNatPhys2011, 
ZiffScience2013, Cho2014Origin, Grassberger2011PRL, Schrenk2011PREGaussian, 
MoreiraPRE2010, Andrade2011PRE, Reis2012PREe}. Finally, after a numerical 
observation \cite{Friedman2009PRL}, there has been given a mathematical proof 
\cite{Riordan2011Science} in favor of the continuous phase transition. The type 
of transition for the explosive percolation in Euclidean space 
\cite{ziff2009PRL, Ziff2010PRE} has not yet been clarified.\\Other variants and 
modifications of percolation models will be discussed in Sec.~\ref{chap3}.

\begin{figure}[h]
\centerline{
\includegraphics[width=0.4\textwidth]{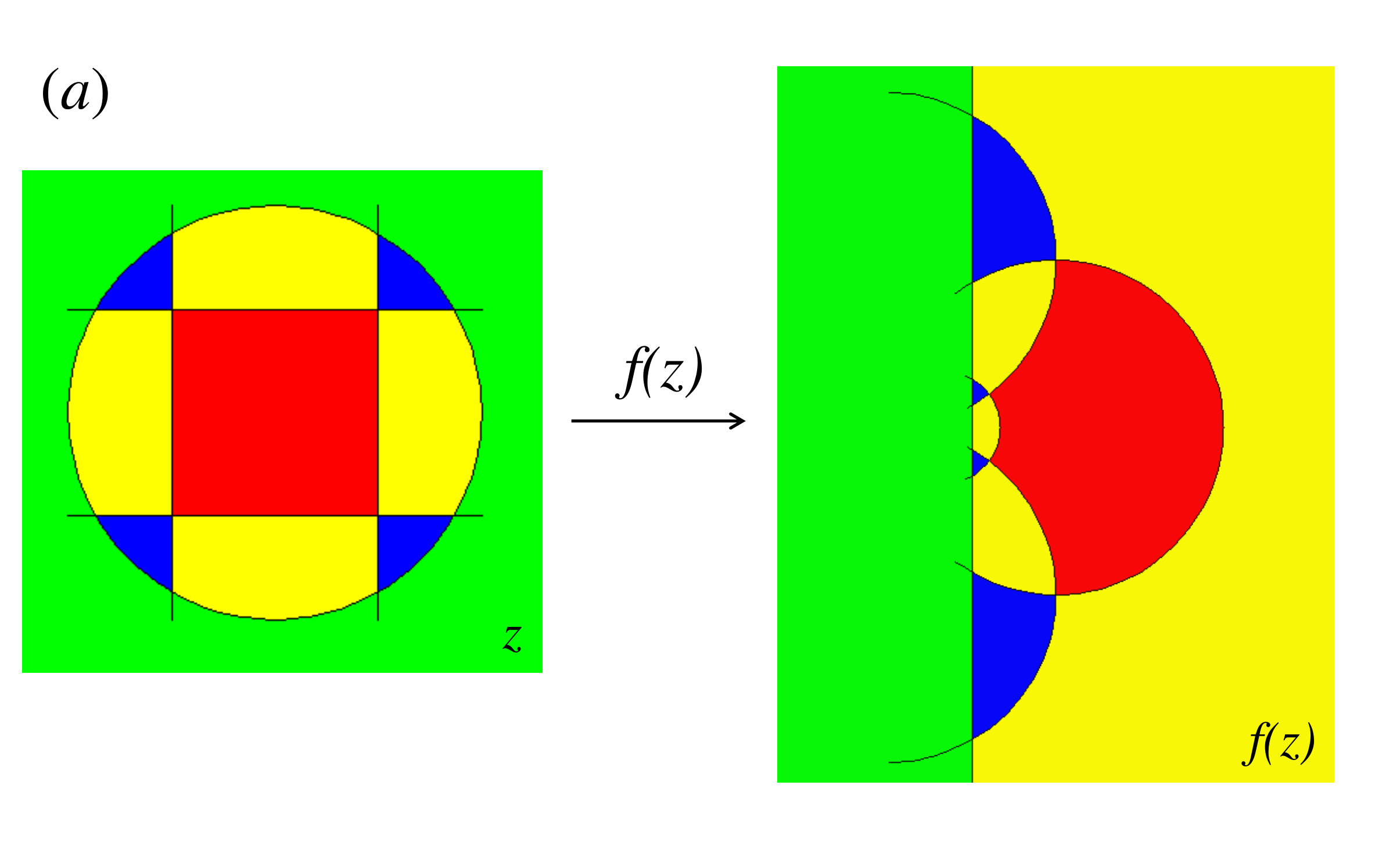}
}
\centerline{
\includegraphics[width=0.5\textwidth]{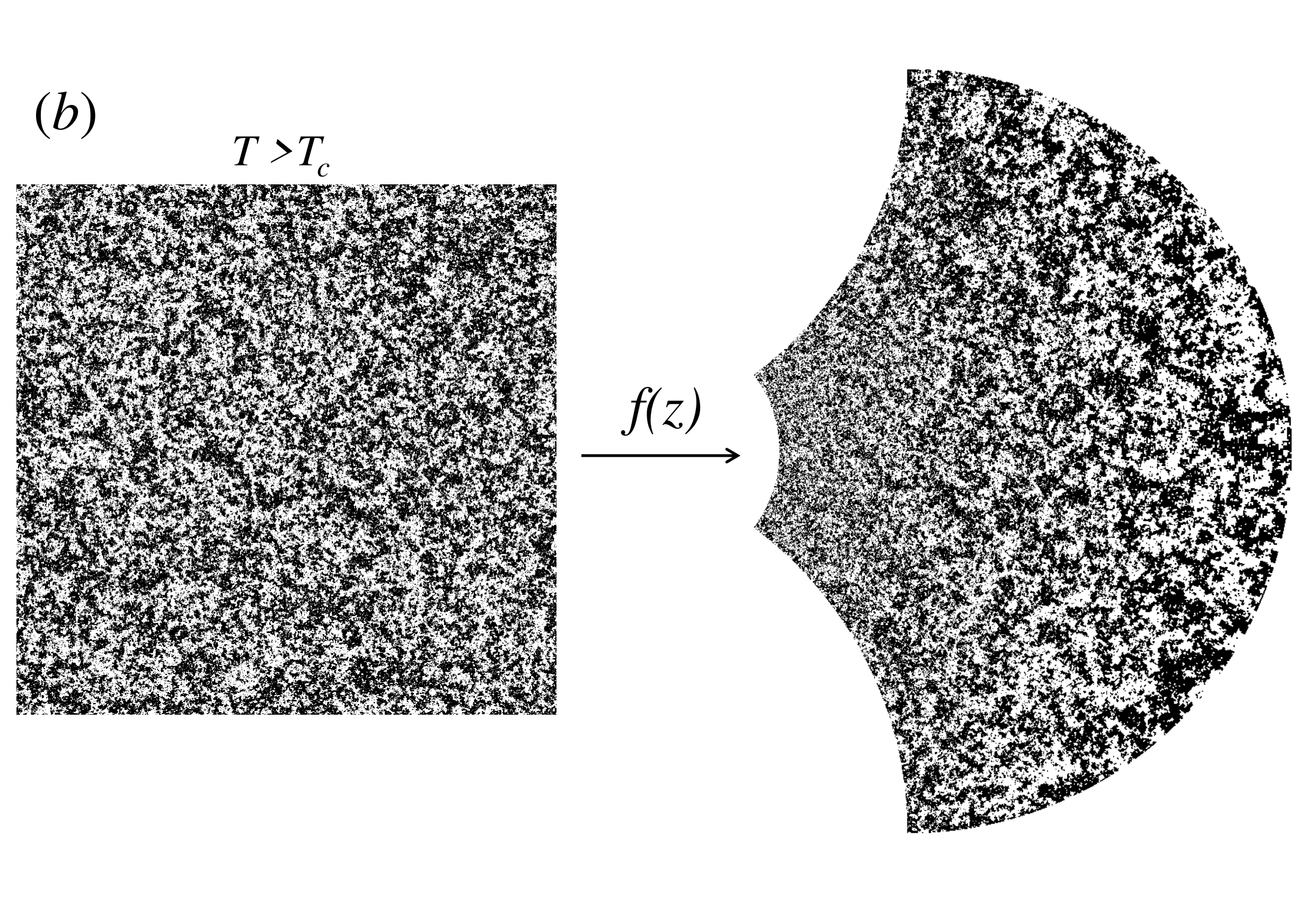}
}
\centerline{
\includegraphics[width=0.5\textwidth]{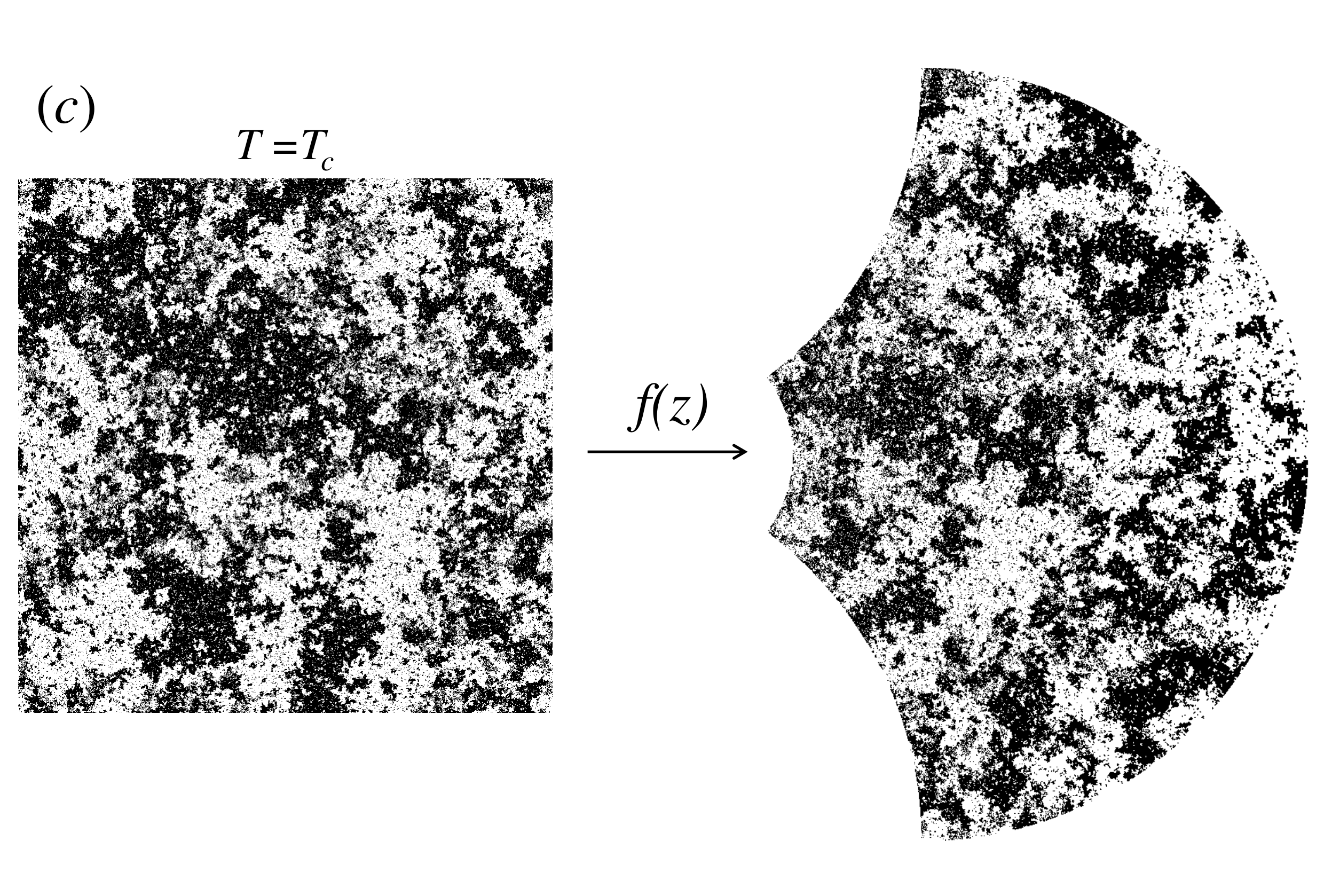}
}
 \caption{(a) Illustration of the transformed image under the special conformal mapping $f(z)=z/(2-z)$. Left panels in (b) and (c) show the homogeneous spin configurations of a 2D Ising model on a square lattice above and at the Curie point, respectively. Under the conformal transformation $f(z)$, the spin configuration above $T_c$ is no longer homogeneous while the one exactly at $T_c$ still is and looks statistically like the original configuration due to its conformal invariant symmetry.}
\label{fig4}
\end{figure}
 
Looking back at Figure~\ref{fig2}, we see that all graphs of different size cross at a single point which is the percolation threshold. This means that at an exactly certain critical occupancy the percolation probability becomes scale invariant. Scale invariance means that if we zoom in or out and look at the system with an arbitrary window size, then the picture still has the same statistical properties with the same resulting physics. This is intimately related to one of the most important methods developed in theoretical physics called \textit{renormalization group}, which studies the behavior of a system under scale transformations \cite{Cardy1998RG}. Renormalization group has been very successful to provide very good approximate (or sometimes exact) values of the critical exponents and thresholds for the percolation models. \\In two dimensions it turns out that an even stronger symmetry holds at the critical point: if we blow up 
different parts of a figure of percolation clusters by different magnification 
factors (as long as angles are preserved) then statistically the picture once again looks the same. 
This property is called conformal invariance, and with this assumption, theoreticians have been able to derive many important properties of critical systems in the past few decades \cite{Cardy2006Nature, Cardy1996Scaling}. In fact the
heuristics behind conformal invariance is a natural generalization of rotation and
scale invariance. Conformal invariance is much more powerful
in two dimensions where it is associated with the
theory of analytic functions of a complex variable. \\Figure~\ref{fig4} 
visualizes how such transformations work. Consider a special conformal 
transformation given by $f(z)=z/(2-z)$ which acts in a way that is shown in 
Fig.~\ref{fig4}(a) on the regions of different color in the complex $z$-plane, 
the map of each domain of specific color in the $z$-plane at left is shown with 
the same corresponding color at right.    
If we have an Ising model on a square lattice and color the spin up (down) 
clusters in black (white), then we see that the figures both above and at the 
Curie point seem to be homogeneous---see left panels in Fig.~\ref{fig4}(b) and~\ref{fig4}(c). However when we look at the transformed spin configurations 
under the conformal mapping $f(z)$, we see that for $T>T_c$ the figure is no 
longer homogeneous, while the one at the critical point still is, and looks 
statistically the same as the original spin configuration at $T_c$ due to its
conformal invariant symmetry.

The emergence of conformal symmetry at the critical point is however more mysterious. This
seems to be a generic feature of criticality but why this happens is not fully understood \cite{Polchinski1988NPB}.
Motivated by numerical experiments \cite{Langlands1992, Langlands1994AMS} in which concluded that crossing probabilities should have a universal scaling limit, which is conformally invariant (a conjecture that is attributed to Michael Aizenman \cite{Langlands1994AMS}), John Cardy used these heuristic ideas in $1992$ to give an explicit formula
that determines the exact values of the crossing probabilities between the opposite
sides of a conformal rectangle filled with a conformally invariant infinitesimal
lattice \cite{Cardy1992Crossing}.  More recently Smirnov rigorously proved \cite{Smirnov2001Proof} that the Cardy’s conjecture holds for the continuum limit of site percolation on a triangular
lattice. But how can one characterize the continuum limit of a lattice model?

The continuum limit of a lattice model is often difficult to be captured mathematically. In this limit the lattice spacing $a$ is sent to zero, and new sites are constantly added to the lattice during
this contraction so that the lattice does not eventually vanish but continues to fill
the original domain that it occupied \cite{Flores2012PhD}. The continuum limits 
of most lattice models are believed to converge to the quantum field theories 
(QFTs). At the critical point, since the correlation length $\xi$ is infinite 
($\xi\gg a$)(Fig.~\ref{fig3}), the lattice models must be invariant under scale 
transformations. This property along with the invariance under translations and 
rotations imply (under broad conditions \cite{Polchinski1988NPB}) the conformal 
invariance, and indeed, suggest that the continuum limit of critical lattice 
models should be given by conformal field theories (CFTs). Furthermore, only 
certain CFTs, usually the minimal models, have been observed to possess the 
right structure to describe a critical lattice model in two dimensions. Due to 
the relatively few number of such theories, models with the same macroscopic 
but different microscopic properties are presumed to have identical continuum 
limits which 
correspond to the same CFT characterized by the value of the central charge $c$. This is a restatement of the notion of \textit{universality} discussed earlier. The critical percolation and Ising models are famous examples of CFTs with central charge $c=0$ and $c=1/2$, respectively. 

A relatively new method to describe the continuum limit of the critical lattice models is called stochastic L\"{o}wner evolution (SLE$_\kappa$) invented by Oded Schramm around 2000 \cite{Schramm2001SLE}---to review SLE see \cite{Cardy2005SLE} and \cite{Bernard2006SLE}. Six years later, in 2006, Wendelin Werner received the Fields medal for his contributions to the development of SLE and related subjects. Theory of SLE$_\kappa$ is a subject of probability theory that generates planar random curves with conformally invariant probability measures in a domain with a boundary. The diffusivity $0\leq\kappa\leq 8$ is a real parameter that classifies different conformally invariant interfaces. For example, the scaling limit of a percolation cluster boundary (hull) is proven by Smirnov \cite{Smirnov2001Proof} to be given by SLE$_6$. It is also shown in \cite{Smirnov2014Ising} that the interfaces in the planar critical Ising model and its random-cluster representation converge
to SLE$_3$ and SLE$_{16/3}$, respectively. In 2010, Smirnov was awarded the Fields medal for the proof of conformal invariance of percolation and the planar Ising model in statistical physics. SLE has soon found many applications and turned out to describe the vorticity lines in turbulence \cite{Bernard2006Nature, Bernard2007PRL}, domain walls of spin glasses
\cite{Amoruso2006PRL, Bernard2007PRB, Davatolhagh2012JSTAT}, the nodal lines of random wave functions \cite{Keating2006PRL, Bogomolny2007JPA}, the iso-height lines of random grown surfaces \cite{Saberi2008PRL, Saberi2008PRE1, Abraham2009EPL, Saberi2009PRE0, Moriconi2010PRE, Saberi2010PRE0}, the avalanche lines in sandpile models \cite{Saberi2009PRE1} and the coastlines and watersheds on Earth \cite{Daryaei2012PRL, AbbasAhmed2012, Boffetta2008GRL}. Among which, SLE could provide quite unexpected connections between some features of interacting systems and ordinary uncorrelated percolation \cite{Bernard2006Nature, Keating2006PRL}.

In 1969, Fortuin and Kasteleyn (FK) \cite{Fortuin1969, Fortuin1972, FK11972, FK21972} found an
interesting mapping between the $q-$state Potts model, which includes the Ising 
model for $q=2$, and a correlated bond-percolation model called the 
random-cluster model. This yielded a geometric representation of the partition 
function for Potts models in terms of the statistics of the random clusters.  
The uncorrelated bond percolation model itself can be recovered by taking the 
limit $q\rightarrow 1$ in the FK formalism. This representation has also become 
a key point to derive many analytical results in percolation 
\cite{Cardy1998RG}. Swendsen and Wang \cite{Swendsen1987PRL}, and then Wolff 
\cite{Wolff1989PRL}, have exploited this mapping to devise extraordinarily 
efficient Monte Carlo algorithms based on nonlocal cluster update for Potts 
models having far less critical slowing down than the
standard single-spin-update algorithms. It turned out that for those values of $q$ for which the model undergoes a continuous phase transition, the percolation of FK clusters
occurs exactly at the critical temperature and their fractal structure encodes the complete critical behavior. It can be shown that there is a one-to-one correspondence between different thermodynamic quantities and their geometric counterparts based on the statistical and fractal properties of FK clusters.

A major breakthrough in statistical physics was the exact solution of the Ising model in two dimensions \cite{Onsager1944}. Onsager gave in 1944 a complete solution of the problem in zero external magnetic field. But in three dimensions, Istrail has shown \cite{Istrail2000NP} that essentially all versions of the Ising model are computationally intractable across lattices and thus the 3D Ising model, in its full, is NP-complete. We will show that is possible to map, at least, the criticality of a 3D Ising model onto a 2D cross section of the model. This mapping provides a dimensional reduction in the geometrical interpretation of the 3D Ising model. Loosely speaking, for the Ising model on square lattice, there exists an alternative description of the partition function as a sum over all curves surrounding the geometric spin clusters (rather than the FK clusters) weighted by their length. The continuum limit of the model in this case is shown to be well-defined and is given by the theory of free Majorana fermions \cite{Distler1992NPB}. Thus our attempt to recast the 2D Ising model as a theory of immersed curves seems successful. But for an Ising model on a cubic lattice, the boundaries between geometric spin clusters are closed surfaces, not curves. It is, however, by no means clear how to take the continuum limit of this lattice surface theory \cite{Dotsenko95NPB, Distler1992NPB, Dotsenko1987NPB3D, Sedrakyan1993PLB1, Sedrakyan1993PLB2}. What we will suggest in Sec.~\ref{chap4}, is to replace the lattice surface theory in 3D criticality, by a theory of immersed curves on a 2D cross section of the original lattice.

\begin{figure}[h]
\centerline{
\includegraphics[width=0.44\textwidth]{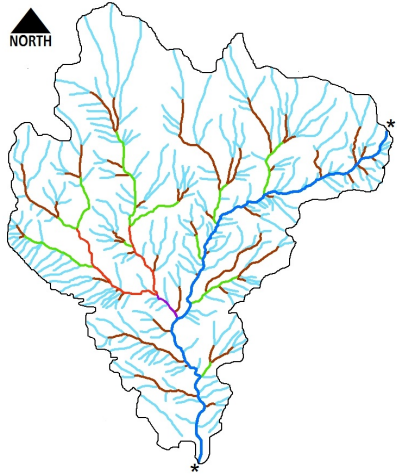}
}
 \caption{Fractal structure of a natural watershed at the north of Tehran, Darabad. The main watershed is the (blue) curve connecting the two points marked by $\star$.}
\label{fig5}
\end{figure}

Percolation theory also provides a suitable platform to study the properties of real and artificial landscapes. A landscape is a height profile usually defined on a square lattice where each cell’s
elevation value at position \textbf{x} represents the average elevation over 
the entire footprint of the cell (site). Now imagine that the water is dripping 
uniformly over the landscape and fills it from the valleys to the mountains, 
letting the water flow out through the open boundaries. During the raining, 
watershed lines may also form which divide the landscape into different 
drainage basins---see also Fig.~\ref{fig5}. Watersheds play a fundamental role 
in geomorphology in e.g., water management \cite{Vorosmarty1998JH} and 
landslide and flood prevention \cite{Lee2006JAWRA}. For a given landscape 
represented as a digital elevation map (DEM), it is possible to determine the 
watershed lines based on the iterative
application of invasion percolation \cite{Fehr2009Watershed}. It is found \cite{Fehr2012WatershedPRE} that the main watershed line generated on random uncorrelated landscapes are self-similar with fractal dimension $d_f= 1.2168\pm0.0005$.
It has been also shown that a watershed lines is statistically compatible with the family of conformally invariant SLE$_\kappa$ curves with $\kappa=1.734\pm0.005$ \cite{Daryaei2012PRL}. Watersheds and the shortest paths on critical percolation clusters with $\kappa=1.04\pm 0.02$ \cite{Herrmann2014Shortest}, are examples of unusual universality classes. Watersheds can also be defined in higher dimensions \cite{Schrenk2012SciRep}. \\ By raising the water level through the landscape, different lakes are gradually forming and start merging to each other to form larger and larger lakes. It is likely to expect that at a certain water level, a percolation transition of the lakes would happen to form a giant lake so that it touches the borders and the water can indeed flow out of the landscape. This problem is closely related to the retention capacity of the landscapes addressed in \cite{Baek2012PRE, SchrenkZiff, Knecht2012ziff}. Whether the percolation transition is critical or not, is strongly dependent on the spatial behavior of the correlations between the height variables \cite{Schmittbuhl1993JPA}. For self-affine surfaces with positive Hurst exponent $H$ \cite{Sahimi1998PhysRep, Sahimi1994JPI, Sahimi1995AIChE, Sahimi1996PRE, Sahimi2000ibid} where the correlation behaves like $\sim (1-r^{2H})$ with the spatial distance $r$, there will not be a genuine percolation transition, while for a long-range correlated surface in which the correlation decays
with the distance as $\sim r^{-2H}$ with $H>0$, the percolation transition is critical \cite{SaberiAPL} and corresponding critical exponents change with $H$ \cite{Sandler2004PRB, Schrenk2013PRE, Weinrib1983PRB, Janke2004PRB}. The fractal dimension of the watersheds \cite{Fehr2012WatershedPRE, Fehr2011PRLWatersheds} as well as various geometric features are dependent on $H$ \cite{Kalda2008EPL, Kondev1995PRL, Kondev2000PRE, Schwartz2001PRL, Mandre2011EPJB}. It has been verified numerically that the duality relation, as a characteristic property of conformally invariant fractals \cite{Duplantier2000PRL}, also holds for the perimeter of the largest cluster in the full range of Hurst exponents \cite{Schrenk2013PRE, PhDthesisScrenk}.

For real landscapes, in addition, percolation theory provides an interesting description for the global topography of Earth. It is found in \cite{Saberi2013PRL} that a percolation transition occurs on Earth's topography in which the present mean sea level is automatically singled out as a critical level in the model. This finding elucidates the origins of the appearance of ubiquitous scaling relations observed in the
various terrestrial features on Earth. This transition is shown to be accompanied by a continental aggregation which sheds light on the possibility of the important role played by water during the long-range topographic evolutions. The criticality of the current sea level also justifies the appearance of the scale (and conformal) invariant
features on Earth, e.g., the fractal rocky coastlines \cite{Boffetta2008GRL}, with an intriguing coincidence of
the dominant $4/3$ fractal dimension in the critical model. The geometrical irregularity of the shorelines actually helps damping the sea waves and decreasing the average wave amplitude. A simple model is accordingly presented in \cite{Sapoval2003}, which produces a stationary artificial shoreline related to the percolation geometry.
A practical application of the discovery of the conformal invariance in the statistical properties of the shorelines is that it allows one to analytically predict
the highly intermittent spatial distribution of the flux of pollutant diffusing ashore \cite{Boffetta2008GRL}.

This review paper is organized as follows. Section~\ref{chap2} presents additional basic properties of percolation model. Formulation of the fractal structure and critical properties of the model is given based on the scaling theory. Section~\ref{chap3} describes different modifications and variants of percolation models which are mostly developed recently. In Section~\ref{chap4} we focus on the geometrical properties of the percolation model in two dimensions. The theory of SLE and its analytical consequences for the percolation model are briefly reviewed. We also show that how one can map the criticality of a 3D Ising model onto a 2D cross section of the original model. Section~\ref{chap5} outlines the statistical properties and the modeling of the seashores and watersheds. We explain how percolation theory can describe the origins of the appearance of various fractal patterns on Earth. Finally, we briefly summarize our results in Section~\ref{conclusion}.

\section{Basic properties of the percolation model}
\label{chap2}

As we saw in the Introduction, percolation theory is concerned with the clustering properties of identical objects which are
randomly and uniformly distributed through space with an occupation probability $p$. It is so simply defined
yet so full of fascinating results. Although it is purely geometrical in 
nature, it embodies many of the important concepts of critical phenomena. In 
order to formulate the model, let us now define a bunch of geometric 
observables. For now we consider a system of infinite lattice size and then in 
Section~\ref{chap2.3}, we will address the situation of finite size effects. 
For a given value of occupancy $p$, the nature of percolation is related to the 
properties of the occupied clusters. Two objects (e.g., occupied sites or 
bonds) belong to the same cluster if they are linked by a path of 
nearest-neighbor bonds joining them (this definition is slightly different for site and bond percolation). The connectedness is the essential 
characteristic of the percolation model denoted by the spanning probability 
$P_s$:  In the limit of an infinite lattice there exists a well-defined 
threshold probability $p_c$ above which there suddenly emerges an infinitely 
large cluster that spans the system (Fig.~\ref{fig2}). Therefore, for $p>p_c$ 
there exists almost surely (i.e., $P_s=1$) one infinite cluster of strength 
$P_\infty(p)$ which denotes the probability for a given site
to belong to the infinite cluster. All other clusters have finite size $s$ at any arbitrary $p$, described by the cluster size distribution $n_s(p)$ i.e., the number of finite clusters of $s$ connected sites, per lattice site. The probability that an arbitrary site belongs to a finite cluster of size $s$ is then given by $sn_s( p)$. Thus, the sum of all the probabilities that a given site belongs to either
a finite size cluster or the infinite cluster must equal to
$p$, i.e., \begin{equation}\label{eq1}
\sum_{s} sn_s(p) + P_\infty(p)=p.
\end{equation}
In the subcritical region, $p<p_c$, $P_\infty(p)$ is identically zero, and in the supercritical case $p>p_c$, $P_\infty(p)$ is positive. Consequently, in the vicinity
of the percolation threshold, the function $P_\infty(p)$ is nonanalytic 
(Fig.~\ref{fig3}).

\begin{figure}[h]
\centerline{
\includegraphics[width=0.6\textwidth]{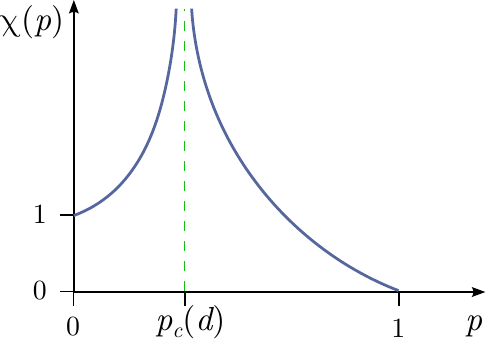}
}
 \caption{Mean cluster size $\chi(p)$ as a function of the occupancy $p$.}
\label{fig6}
\end{figure}

Now we need to define a quantity to measure the cluster size.
Intuitively, at a low value of $p$ cluster sizes are small and increase with $p$ until the threshold where the spanning cluster dominates and is infinite in
size, and thus the cluster size must diverge. Above the threshold we should
remove the infinite cluster from our calculations otherwise it will always dominate. As the clusters get absorbed into the spanning cluster
the typical size of those left goes back down again. Therefore, we should have a cluster size which
increases, diverges at the threshold and then decreases again.\\By Bayes' theorem, the probability a site belongs to a cluster of finite mass $s$ given that the site is occupied is \begin{equation}
w_s(p)=\dfrac{sn_s(p)}{\sum_s sn_s(p)}.
\end{equation}\label{eq2}
Thus we can define the mean cluster size $\chi(p)$ as follows,
\begin{equation}\label{eq3}
\chi(p)=\sum_s sw_s(p)= \dfrac{\sum_s s^2n_s(p)}{\sum_s sn_s(p)}.
\end{equation}
This definition (\ref{eq3}) however, does not have information about the structure of the clusters e.g., their compactness and spatial extent. For each cluster of mass $s$ one can instead measure its radius of gyration $R_{g,s}$, defined by $R^2_{g,s}=(1/2s^2)\sum_{i,j}(r_i-r_j)^2$, where the sum is over pairs of
points on the cluster. For a given cluster of size $s$, the smaller radius of 
gyration is indicative of its higher compactness and lower spatial extent.\\The 
probability to find a finite cluster of large size $s$ at a given point 
decreases exponentially with $s$ in the subcritical regime 
\cite{Grimmett1999BookCambridge}. More precisely, there exists $\kappa(p)>0$, 
so that $\kappa(p)\rightarrow\infty$ as $p\rightarrow0$ and $\kappa(p=p_c)=0$ 
such that \begin{equation}\label{sizedecay}
w_s(p)\approx e^{-\kappa(p) s}, \hspace{1cm} s\longrightarrow\infty.
\end{equation}
 It can also be shown that in the supercritical regime, the tail of the finite cluster size distribution has a rather smoother decaying form. In other words, there exist functions $\kappa_1(p)$ and $\kappa_2(p)$, satisfying $0<\kappa_2(p)\leq\kappa_1(p)<\infty$, such that \cite{Grimmett1999BookCambridge} \begin{equation}\label{decaysuper}
\exp\big(-\kappa_1(p) s^{(d-1)/d}\big) \leq w_s(p) \leq \exp\big(-\kappa_2(p) s^{(d-1)/d}\big).
\end{equation}
Note that the power $s^{(d-1)/d}$ is the order of the surface area of the 
sphere in $d$ dimensions with volume $s$. This implies that the clusters are compact in the supercritical region~\cite{Djordjevic1987}.\\At the critical point, this 
probability has a power-law decaying form of $w_s(p_c)\approx s^{-1-1/\delta}$, 
with some critical exponent $\delta=\delta(d)>0$---see also Fig.~\ref{fig6}.

We can also define another length scale which is different from the average radius of clusters, i.e., the correlation length $\xi$, defined by the two-point correlation function $g_c(r)$. This is the
probability that if one point is in a finite cluster then another point a distance $r$ away is in
the same cluster. This function then typically has an exponential decay given by a correlation
length $\xi$ \begin{equation}\label{gc}
g_c(r)\sim e^{-r/\xi},\hspace{1cm} r\longrightarrow\infty.
\end{equation} The correlation length is a characteristic size of the cluster distribution which yields a maximum size above which the clusters are exponentially
scarce. It is also the
upper bound of the scaling region where percolation clusters
have a self-similar behavior. Therefore we may define the correlation length $\xi$ as an average distance of two points belonging to the same cluster \begin{equation}
\xi^2=\dfrac{\sum_r r^2g_c(r)}{\sum_r g_c(r)}.
\end{equation}
For a given cluster of mass $s$, one may replace $r^2$ in the above summation by the average squared distance between two cluster points i.e., $2R^2_{g,s}$. Moreover, with the probability $sn_s$, a point belongs to an $s$-cluster, and since it is then connected to $s$ sites, one may also replace $g_c(r)$ by $s^2n_s$, giving rise to the following relation for the squared correlation length \footnote{The mean-square distance between two sites on an $s$-cluster is
related to $R_{g,s}$, $(1/s^2)\sum_{i,j}(r_i-r_j)^2=2R^2_{g,s}$, where the factor of $2$ comes from counting each pair twice. The correlation length $\xi$
can be defined as an average distance between two cluster sites.
Whereas $2R^2_{g,s}$ is the mean-square distance between two sites on an $s$-cluster, $\xi^2$ is this same distance averaged over all finite sizes $s$.
The probability that a site belongs to $s$-cluster is $sn_s$.
There are $s$ sites in each $s$-cluster. One can thus weight the average
of $2R^2_{g,s}$ by $s\cdot sn_s$, to obtain the squared correlation length Eq.~(\ref{xi2}).} 
\begin{equation}\label{xi2}
\xi^2(p)= \dfrac{\sum_s 2R^2_{g,s}s^2n_s(p)}{\sum_s s^2n_s(p)}.
\end{equation}
The above definitions of different characteristic observables are valid in all dimensions $d$.

\subsection{Percolation in \textit{d}-dimensions}

\subsubsection{Percolation on $\mathbb{Z}^d$}

Let us consider the percolation problem on a hypercubic lattice $\mathbb{Z}^d$ 
in $d$-dimensions. In $d=1$, it is a trivial task to find that the critical 
threshold should be $p_c=1$ both for bond or site percolation models. For 
$d>1$, there exists a critical threshold $0<p_c<1$ below which all open 
clusters are finite and there is, almost surely, no infinite open cluster, and 
above $p_c$ there exists an infinite open cluster with probability 1. It is 
rigorously known that no infinite open cluster exists at $p=p_c$ for $d=2$ and 
$d\geq19$ \cite{Hara1994Meanfield}. For other dimensions it is also conjectured 
to be held but its proof is viewed as an important open mathematical problem in 
the field. It is also known that for an infinite connected graph with maximum 
finite vertex degree $\Delta$, the bond and site critical thresholds i.e., 
$p_c^{b}$ and $p_c^{s}$, respectively, satisfy the following inequality 
\cite{Grimmett1999BookCambridge} \begin{equation}
\dfrac{1}{\Delta-1}\leq p_c^b\leq p_c^s\leq 1-(1-p_c^b)^\Delta.
\end{equation}   
In particular, $p_c^b\leq p_c^s$ where the strict inequality holds for a broad family of graphs.
It can also be shown that for percolation on $\mathbb{Z}^d$ the percolation probability always satisfies $P^{(d+1)}_\infty(p)\geq P^{(d)}_\infty(p)$ for all $p$ and $d$, and consequently we have $p_c(d+1)\leq p_c(d)$ \cite{SteifMiniCourse, Benjamini1996Percolation}.

In order to further figure out the global feature of percolation, a natural question would be concerned about the number of possible infinite clusters which can coexist. It was shown by Newman and
Schulman \cite{Newman1981Number} that for periodic graphs at any arbitrary $p$, exactly
one of the following three situations prevails with probability 1: The number of infinite open clusters can be either 0, 1 or $\infty$. It has been proved in \cite{Aizenman1987Uniqueness} that
the third situation is impossible on $\mathbb{Z}^d$. It has also been proved in \cite{Burton1989Unique} that
there cannot be infinitely many infinite open clusters
on any amenable graph~\footnote{For a finite set of vertices $V$ whose edge boundary is denoted by $\partial_E V=$\{($u,v$):$u\in V$, $v \notin V$\}, the notion of amenability is related to whether the size of $\partial_E V$ is of equal order as
that of $V$, or is much smaller. Denoting the \textit{Cheeger} constant of a graph $G$ by $h(G)=\underset{V\subset V(G):|V|<\infty}\inf \frac{|\partial_E V|}{|V|}$, that is the minimal ratio of boundary to bulk
of its nontrivial subgraphs, a graph is called amenable when $h(G)=0$, and non-amenable otherwise. The simplest example of a non-amenable graph is the Bethe lattice with $z\ge 3$ for which the \textit{Cheeger} constant is $h=z-2$~\cite{Benjamini1996Percolation, Hofstad2010}. It is also shown \cite{Benjamini1996Percolation} that $p_c(G)\le 1/(h(G)+1)$, so that for every non-amenable graph $p_c(G)<1$.   }. Nevertheless there are some graphs, such as regular trees, on which infinitely many infinite clusters can coexist \cite{Grimmett1990Newman}.

\subsubsection{Percolation on Bethe lattices}

\begin{figure}[t]
\centerline{
\includegraphics[width=0.6\textwidth]{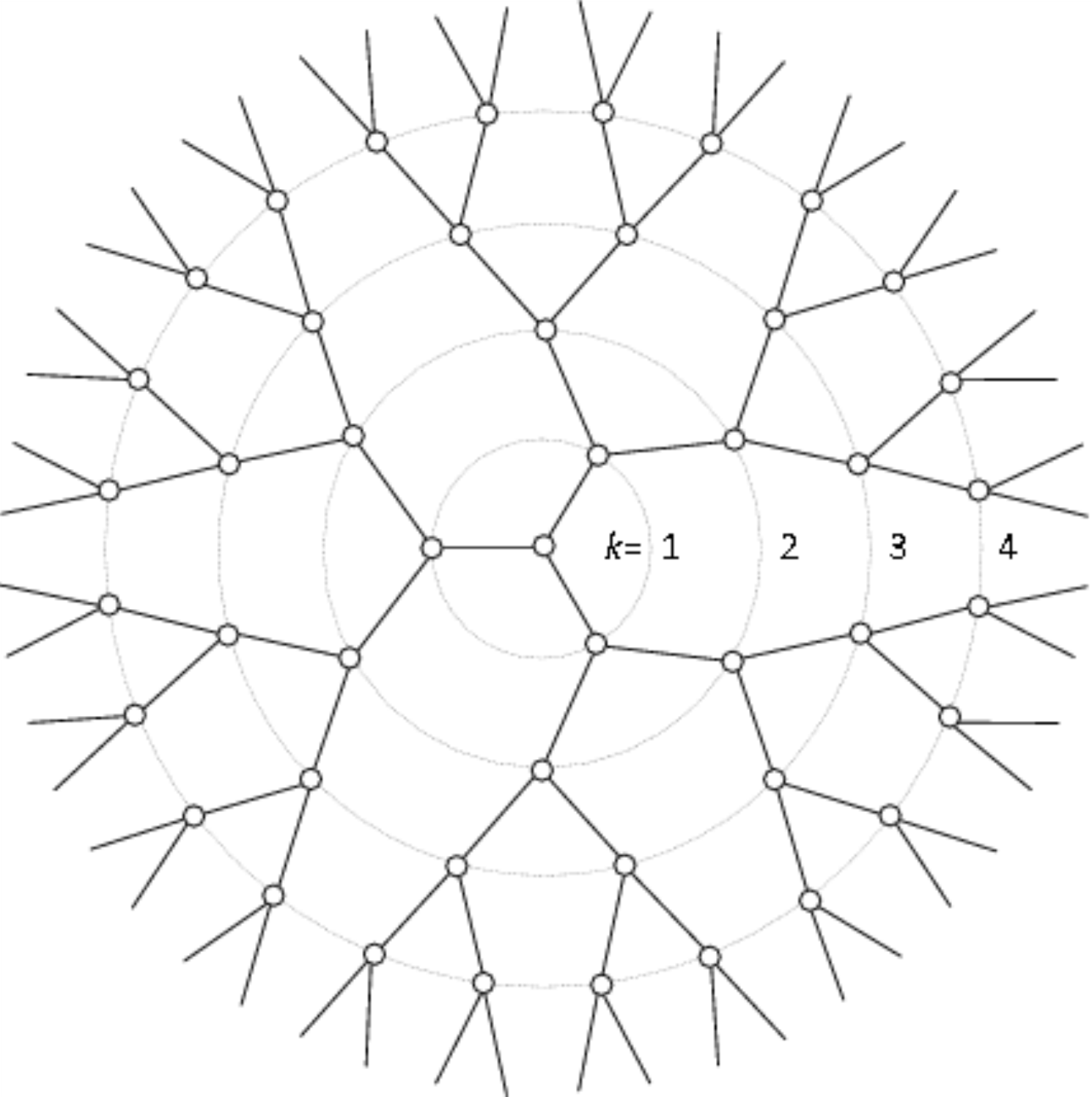}
}
 \caption{Part of a Bethe lattice with coordination number $z=3$ \cite{saberiBethelattice}. The lattice sites are represented by open small circles at different shells $k=0, 1, 2, \cdots$.}
\label{fig7}
\end{figure}

Due to its distinctive topological structure, several statistical
models even with interactions defined on the Bethe lattice
\cite{Bethe1935} are exactly solvable and computationally inexpensive
\cite{Baxter1982EXACTLY}. Various systems including magnetic models
\cite{Bethe1935} and percolation \cite{Thorpe1982Excitations, Sahimi2003Heterogeneous}, have been studied on
the Bethe lattice whose analytic results gave important physical
insights to subsequent developments of the corresponding research
fields. The Bethe lattice is defined as a graph of infinite points
each connected to $z$ neighbors (the coordination number) such that
no closed loops exist in the geometry---see Fig.~\ref{fig7}. A
finite type of the graph with boundary is also known as a Cayley
tree and possesses the features of both one and infinite dimensions:
since $N_k$, the total number of sites in a Bethe lattice with $k$
shells, is given as $N_k=[z(z-1)^k-2]/(z-2)$, the lattice dimension
defined by $d=\lim_{k\rightarrow\infty}[\ln N_k/\ln k]$ is infinite.
It is therefore often mentioned in the literature that the Bethe
lattice describes the infinite-dimensional limit of a hypercubic
lattice $\mathbb{Z}^d$ (It is also clear that a Bethe lattice with $z=2$ is isomorphic 
to the positive 1D lattice $\mathbb{Z}^+$). The Bethe lattice is indeed a very important substrate or medium on
which the mean-field theories for various physical models become exact.
\\As the number of shells grows the number of sites in the surface, or
the last shell, grows exponentially $z(z-1)^{k-1}$. Therefore, as
$k$ tends to infinity, the proportion of surface
sites tends to $(z-2)/(z-1)$. By surface boundary we mean the set of
sites of coordination number unity, the interior sites all have a
coordination number $z$. Thus the vertices of a Bethe lattice can be
grouped into shells as functions of the distances $k$ from the
central vertex. 

It is possible to show that the critical threshold for the Bethe lattice is $p_c=1/(z-1)$ for any $z\geq3$. Moreover, $P_\infty=0$ indicates the subcritical region $p<p_c$, while $P_\infty>0$ indicates the supercritical region $p>p_c$ in which $P_\infty(p)$ is strictly increasing function of $p$. It is also known that at the critical point $P_\infty(p=p_c)=0$, meaning that there is no infinite cluster almost surely at $p_c$. 

Since the Bethe lattice has a tree structure, the only way to connect a point sitting at $k$th shell to the origin is by a path of $k$ edges. Thus the two-point correlation function is given by $g_c(k)=p^k$, which decays exponentially fast as $k\rightarrow\infty$ for all $p<1$. Comparing this with relation (\ref{gc}), it turns out that the correlation length should be $\xi(p)\sim -1/\ln p$. Therefore, unlike for the $\mathbb{Z}^d$ lattice, the correlation length on the Bethe lattice is finite for all $0<p<1$.
\\The mean cluster size on the Bethe lattice can also be explicitly computed as follows,
\begin{equation}
\chi(p)=
\left\{
\begin{array}{l c}
\big(1-(z-1)p\big)^{-1} & \quad  p< p_c,
\\
\infty & \quad \mbox{otherwise}.
\end{array} \right.
\end{equation}
In the next section we will see how $\chi(p)\rightarrow\infty$ as $p$ approaches $p_c$ from the left.

A much wider class of interesting graphs is that of Cayley
graphs (which includes also Cayley trees as a particular kind) of infinite, finitely generated groups.
There, it has been shown \cite{Schonmann1999Percolation} that the number $\mathcal{N}_\infty$ of infinite clusters satisfies

\begin{equation}
\mathcal{N}_\infty=
\left\{
\begin{array}{c c c}
 0 &  \mbox{if} &  p\in [0,p_c),
\\
\infty &  \mbox{if} &  \ \ p\in (p_c,p_u),
\\
1 &  \mbox{if} &  p\in (p_u,1].
\end{array}  \right.
\end{equation}
The parameter space is thus split into three qualitative intervals separated by two critical values $p_c$ and $p_u$. Some of intervals may be however degenerate or empty e.g., for $\mathbb{Z}^d$ we have $p_c=p_u$, and for trees we have $p_u=1 $.

\subsubsection{Percolation on random graphs and networks}\label{randomgrapghs}

The field of random graphs was started in 1959 by Erd\H{o}s and R\'{e}nyi \cite{Erdoes1959IPMD, Erdoes1960MAMK, Erdoes1961BIIS, Erdoes1961AMASH} whose work instigated a great amount
of research in the field \cite{BookRandomgraph2001, JansonRandomgraph2000}. Random graphs have been extensively used as a probabilistic approach to study complex networks \cite{Hofstad2014Randomgraph}. Many real-world complex networks such as the Internet \cite{Barabasi1999Science},
social networks \cite{Watts1998Nature}, disease modeling \cite{Durrett2007Book} etc., share similar features e.g., they are \textit{large}, \textit{sparse}, \textit{scale-free}, \textit{small worlds} and \textit{highly clustered} \cite{WattsSmallworld1999}. Consider a graph $G_n$, with $n$ number of vertices. Denote the proportion of vertices with degree $k$ in $G_n$ by random variable $P^{(n)}_k$. We first call a random graph process sparse when $\lim_{n\rightarrow\infty}P^{(n)}_k=p_k $, for some deterministic limiting probability distribution $\{p_k\}_{k=0}^\infty$. Also, since $\{p_k\}_{k=0}^\infty$ sums up to one, for large $n$, most of
the vertices have a bounded degree, which explains the phrase sparse random graphs.
We further call a random graph process scale-free with exponent $\tau$ when
it is sparse and when 
\begin{equation}
\lim_{k\rightarrow\infty}\dfrac{\log p_k}{\log (1/k)}=\tau
\end{equation}
exists. This means that the number $N_k$ of vertices with degree $k$ is proportional to an inverse
power of $k$, i.e., $N_k\sim c_n k^{-\tau}$ where $c_n$
is some normalizing constant. The requirement $\sum_k N_k=n$ then makes it reasonable to assume that $\tau>1$. \\Let the random variable $H_n$ denote the typical distance between two uniformly chosen connected vertices in $G_n$. Then, we say that the random graph
process is a small world when there exists a constant $K$ such that with probability one, $H_n\leq K\log n$, in the limit of $n\rightarrow\infty$. For ultra-small world random graphs we have almost surely $H_n\leq K\log\log n$.

The simplest imaginable random graph is the Erd\H{o}s-R\'{e}nyi random graph ER$_n(p)$, which arises by taking
$n$ vertices, and placing an edge independently between any pair of distinct vertices with some fixed
probability $p$. This random graph is shown to exhibit a percolation phase transition in the size of the
maximal component, as well as in the connectivity of the arising random graph. The phase transition in ER$_n(p)$, refers to a sharp transition in the largest connected component and serves as the mean-field case of percolation.

The degree of a vertex in ER$_n(p)$ has a binomial distribution with
parameters $n$ and $p=\lambda/n$. It is well known that for $n$ is large, the proportion of vertices with degree $k$ converges in probability to the Poisson distribution with parameter $\lambda$ i.e., \begin{equation}
p_k(\lambda)\longrightarrow e^{-\lambda} \dfrac{\lambda^k}{k!},\quad \text{as} \quad n\longrightarrow\infty, \qquad\qquad k=0, 1, 2, \cdots.
\end{equation}
In particular, $p_0=e^{-\lambda}$ is the fraction of isolated vertices. The control parameter $\lambda$ may thus be seen as the average degree of nodes.\\
Therefore, ER$_n(p)$ is a sparse random graph process but not scale-free, since clearly it does not have a power-law degree sequence. However, in order to adapt the random graph to complex networks, we
can make these degrees scale-free in a generalized random graph by taking the parameter $\lambda$ to be a random variable with a power-law distribution~\cite{Britton2006Generatin}.\\
Erd\H{o}s and R\'{e}nyi have shown \cite{Erdoes1960MAMK} that there is a critical value $\lambda_c=1$ below which ER$_n(p)$ random graph has almost surely no connected components of size larger than $\mathcal{O}(\log n)$.  At $\lambda=\lambda_c$, the graph has a largest component of size $\mathcal{O}(n^{2/3})$, while for $\lambda>\lambda_c$, a drastic separation takes place between the largest cluster and all other smaller components and there appears a unique giant component of $\mathcal{O}(n)$ containing a positive fraction of the vertices.\\ The average fraction $\mathcal{N}(\lambda)$ of clusters (number of clusters per vertex) can be shown~\cite{Monasson2013lecture} to be given by \begin{equation}
\mathcal{N}(\lambda)=-\dfrac{\lambda}{2}(1-P_\infty(\lambda)^2)+(1-P_\infty(\lambda))\big[1-\ln\big(1-P_\infty(\lambda)\big)\big],
\end{equation}
where $P_\infty(\lambda)$ denotes for the fraction of vertices in the largest component. 

As mentioned earlier, the random graphs like ER$_n(p)$ are quite unlike real-world networks, which
often possess power-law or other highly skewed degree distributions. The study of percolation on graphs with completely general degree distribution is, however, presented in \cite{Callaway2000}, giving exact solutions for a variety of cases,
including site percolation, bond percolation, and models in which occupation probabilities depend on
vertex degree.

\subsection{Percolation at and near criticality}

We have learned that the behavior of percolation process depends strongly on whether we are in the subcritical $p<p_c$ or supercritical $p>p_c$ regime. In the former case, all clusters are finite and their size distribution has a tail which decays exponentially---see (\ref{sizedecay}). In the latter supercritical regime, there exists an infinite cluster with probability one and, the size distribution of other finite clusters has a tail which decays slower than exponentially---see (\ref{decaysuper}). At the vicinity of the critical point $p\sim p_c$, however, there occurs an interesting phenomena characterized by a nonanalytic behavior of the order parameter $P_\infty(p)$ along with the divergent asymptotic behavior of the correlation length $\xi(p)$ and the mean cluster size $\chi(p)$ when $p$ approaches $p_c$. Despite lacking their mathematically rigorous justifications, renormalization and scaling theory have made remarkable predictions about the behavior of the percolation problem near and at the critical threshold.

\subsubsection{Scaling hypotheses and upper critical dimension}

According to the scaling hypotheses it is possible to state, in general, the following relation for the number $n_s(p)$ of finite clusters per site  \begin{equation}
n_s(p)\propto s^{-\tau}F\big(c(p)s\big), \qquad s\longrightarrow\infty
\end{equation} where $\tau$ is a free exponent and $F$ is a scaling function. Near the percolation threshold, $c(p)$ is allowed to behave as a general power-law $c(p)\propto |p-p_c|^{1/\sigma}$, where $\sigma$ is another critical exponent. We can consider in general, the $m$th moment of the cluster size distribution defined by $M_m(p)=\sum_s s^m n_s(p)$ with $m\geq1$. The following scaling relations are also conjectured to hold near the percolation threshold with different critical exponents $\beta$, $\gamma$, $\alpha$, $\Delta$ and $\nu$ \begin{subequations}\label{scalinrelations}
\begin{alignat}{2}\label{a}
&\text{percolation strength ($p>p_c$):}\qquad & P_\infty(p)\simeq& 
B(p-p_c)^\beta,\\
&\text{mean cluster size:}\qquad &\chi(p)\simeq&\Gamma^{\pm}|p-p_c|^{-\gamma},\\
&\text{mean cluster number per site:} \qquad &n_c(p)\simeq& A^\pm 
|p-p_c|^{2-\alpha},\label{c}\\
&\text{cluster moments ratio ($m\ge2$):}\qquad 
&\dfrac{M_{m+1}(p)}{M_m(p)}\simeq& 
D^\pm|p-p_c|^{-\Delta},\\
&\text{correlation length:}\qquad &\xi(p)\simeq& f^\pm |p-p_c|^{-\nu},\label{e}
\end{alignat}
\end{subequations}
which also define the critical amplitudes whose superscripts $+$ or $-$ refer 
to $p_c$ being approached from above or below, 
respectively~\footnote{Eq.~(\ref{c}) is valid only for the 
nonanalytic part of $n_c(p)$.}. Universal combinations of these amplitudes 
represent the canonical way of encoding the universal information about the 
approach to criticality \cite{Privman1991book, 1980AharonyPRB}.
While critical exponents can be determined working \textit{at} criticality, amplitude ratios also characterize the scaling region \textit{around} the critical point which carry independent information about the universality class \cite{Viti2010arXiv}. \\The probability for two sites separated by a distance $r$ to belong to the same cluster also has a power-law decaying form $g_c(r)\simeq r^{2-d-\eta}$ for large distances at $p=p_c$, introducing the anomalous dimension $\eta$.

These exponents, however, are not independent of each other but satisfy two sets of scaling and hyperscaling relations. The scaling relations can be easily read as $\gamma=\nu(2-\eta)$, $2-\alpha=\gamma+2\beta=(\tau-1)/\sigma$ and $\beta=\Delta(\tau-2)$. The hyperscaling relation, on the other hand, involves the number $d$ of dimensions $d\nu=2-\alpha$, and believed to be valid only for $d\leq d_c$, where $d_c$ is called \textit{upper critical dimension}. It is believed that when $d\geq d_c$, the percolation process behaves roughly in the same manner as percolation on an infinite regular tree and their critical exponents take on the corresponding values given by mean-field theory:\begin{equation}\nonumber
\alpha=-1,\quad \beta=1,\quad \gamma=1,\quad \tau=\dfrac{5}{2},\quad \delta=2, \quad \Delta=2, \quad \eta=0, \quad \nu=\dfrac{1}{2}. 
\end{equation}
If these values are attained by percolation, then the hyperscaling relation gives $d_c=6$.\\The critical exponents are known exactly only in 2D, and not much is known rigorously in the general case. Critical exponents are \textit{universal} in the sense that they depend only on dimensionality $d$, and not otherwise upon the individual structure of the underlying lattice. Theory of renormalization group (RG) lends support to the hypothesis of universality. 

\subsubsection{Real-space renormalization group}

The Kadanoff picture of RG, called real-space RG, is based on coarse graining and rescaling procedure in which the lattice is iteratively divided into blocks of linear size $b$ and then rescaled. When this scale transformation is iterated many times, RG leads to a certain number of fixed points. The fixed point equation for the rescaling transformation $\xi=\xi/b$ in percolation, has two solutions only: $\xi=0, \infty$. These are associated with the solutions to the fixed point equation in $p$-space, $T_b(p)=p$, that is, $p=0, 1$ and $p_c$, representing the trivially self-similar states of the empty and fully occupied lattice and the nontrivial self-similar state at $p = p_c$, respectively. The critical exponent $\nu$ can then be given by \begin{equation}\nonumber
\nu=\dfrac{\log(b)}{\Bigg(\dfrac{dT_b(p)}{dp}|_{p_c}\Bigg)}\approx \dfrac{\log(b)}{\Bigg(\dfrac{dR_b(p)}{dp}|_{p^\star}\Bigg)},\end{equation} where the rescaling transformation $T_b$ has been substituted with a real-space
renormalisation transformation $R_b$ incorporating coarsening with rescaling. Note that $p_c$ is identified with $p^\star$, the nontrivial solution to the fixed point equation $R_b(p^\star)=p^\star$. The real-space renormalisation transformation $R_b(p)$ is often chosen to be the 
probability of having a spanning cluster (or a majority of sites occupied) in the block.\\ One can therefore summarize the real-space 
renormalisation transformation procedure as follows:
i) Divide the lattice into blocks of linear size $b$.
ii) Replace all sites in a block by a single block of size $b$ occupied with probability $R_b(p)$ according to the coarse graining procedure.
iii) Rescale all length scales by the factor $b$.\\
As critical exponents are determined by the large scale behavior, they are universal and
insensitive to details of lattice structure. However, real-space renormalization gives limited results and does not give the
critical exponents, except in the case of the Bethe lattice.

\subsection{Fractal structure of the critical percolation clusters}
\label{chap2.3}
Scaling theory asserts that whenever system is viewed on length scales smaller than the correlation length $\xi$, it behaves as it does at the threshold. At the critical point, $\xi$ as the only length scale dominating the critical behavior of an infinite lattice, is diverging i.e., $\xi\rightarrow\infty$. Disappearance of this scale at $p=p_c$, is reminiscent of scale invariance which implies the emergence of self-similarity in the geometric feature of the percolation clusters. The fractal properties persist even for $p\neq p_c$ with finite $\xi$, whenever the length scale to investigate the system is less than $\xi$; once it exceeds $\xi$, the geometry becomes Euclidean. Consider a percolation system which is viewed through a hypercubic window of size $L^d$ where $L\ll\xi$ is the linear window size or can be regarded as the size of a finite system. Scale invariance then requires that the mean mass $M$ of the cluster within the window would increase as a power-law with size i.e., $M(\xi,L)\sim L^{d^c_f}$, where $d^c_f$ is the fractal dimension of the cluster. Above $p_c$ on length scales $L\gg\xi$, the infinite cluster can be regarded as a homogeneous object which is composed of many cells of size $\xi$, i.e., $M(\xi,L)\sim\xi^{d^c_f}(L/\xi)^d$. These can be mathematically summarized in terms of the crossover function $\mathfrak{m}$ as follows, \begin{equation}\label{MassScaling}
M(\xi,L)\sim L^{d^c_f}\mathfrak{m}(L/\xi), \quad \mbox{where} \quad \mathfrak{m}(L/\xi)=
\left\{
\begin{array}{c c c}
 \mbox{constant}&  \mbox{for}  & L\ll\xi,
\\
(L/\xi)^{d-d^c_f} &  \mbox{for} & L\gg\xi.
\end{array}  \right.
\end{equation}
The mass $M$, on the other hand, is proportional to $L^d P_\infty$. Equating this with (\ref{MassScaling}) and rewriting in terms of $(p-p_c)$ using  (\ref{a}) and (\ref{e}), yields the scaling relation $d_f^c=d-\beta/\nu$. Since the exponents $\beta$ and $\nu$ are universal, the fractal dimension $d_f^c$ is also universal. The values of $d_f^c$ are known exactly only in 2D and $d\ge d_c=6$ as $d_f^c=91/48$ and $4$, respectively. In other dimensions the estimates exist only by numerical simulations. However the fractal dimension $d_f^c$ by itself is not enough to characterize the percolation cluster. For example, a critical percolation cluster and a pattern of diffusion-limited aggregation (DLA) in 3D, share the same value of fractal dimension $d_f^c\simeq 2.5$, while their fractal structures have completely different features. For its better characterization one can define the shortest path between two cluster points. The shortest path of length $l_{min}$ is shown to be self-similar which satisfies the scaling relation $l_{min}\sim R^{d_f^{min}}$ where $R$ is the linear distance between the two points and, $d_f^{min}$ denotes for the shortest path fractal dimension. The chemical dimension $d^{ch}$ is then defined by $M\sim l_{min}^{d^{ch}}\sim R^{d_f^{min} d^{ch}}$ which means $d_f^c=d_f^{min} d_f^{ch}$. The fractal dimension $d_f^{min}$ is known exactly only for dimensions $d\ge 6$ to be $d_f^{min}=2$. Even in 2D, despite its relevance, the fractal dimension
of the shortest path is among the few critical exponents that are not known
exactly \cite{Grassberger1985JPA, ZhouZiff2012PRE}. Now $d_f^{min}$ distinguishes between percolation cluster and DLA in 3D, by $d_f^{min}\simeq1.38$ and $1$, respectively.\\
A fractal percolation cluster is composed of several other fractal substructures including its perimeter (hull), external perimeter, backbone, dangling ends and red sites (bonds), etc. For instance, the mean number of red bonds $N_r$ varies with $p$ as $N_r(p)\sim (p-p_c)^{-1}$, implying $N_r\sim\xi^{1/\nu}$ which gives the fractal dimension of the red bonds $d_f^r=1/\nu$, valid in all dimensions~\cite{Coniglio1982JPA}.\\ In 2D, the perimeter and the external perimeter have the fractal dimension $d_f^P=7/4$ and $d_f^{EP}=4/3$, respectively, which belong to the family of conformally invariant curves called SLE$_\kappa$, with $\kappa=6$ and $\tilde{\kappa}=8/3$, respectively, satisfying the duality relation $\kappa\tilde{\kappa}=16$.

The fractal dimensions for the percolating cluster at criticality in ER$_n(p)$ random graph and random scale-free networks with $n$ equal to the number of vertices and degree distribution $p_k=c_nk^{-\tau}$, are also reported in \cite{Cohen2004PhisicaA}. There has been shown that the fractal dimension of the spanning cluster is $d_f^c=4$ for $\tau>4$ and $d_f^c=2(\tau-2)/(\tau-3)$ for $3<\tau<4$. Note that the result for $\tau>4$ is in agreement with that of the regular infinite dimensional percolation. As we mentioned earlier in subsection~\ref{randomgrapghs}, on a random network in the well connected regime, the average distance between sites is of the order $\log_{\langle k\rangle}n$, and becomes even smaller
in ultra-small world random graphs. However as discussed above, for $d\ge 6$ we have $d_f^c=4$ and $d_f^{min}=2$, consequently $d^{ch}=2$. Therefore the average chemical distance $l_{min}$ between pairs of sites on the spanning cluster for random graphs and scale-free networks with $\tau>4$ at criticality behaves as $l_{min}\sim \sqrt{M}$, i.e., the distances become much larger at criticality.

\section{Variants of percolation
}
\label{chap3}

In the previous section, we studied the standard version of percolation i.e, Bernoulli percolation, in which all random occupations of bonds or sites take place irreversibly and independently on either a Euclidean lattice or a random graph. Over the past decades and years, numerous variations and modifications of the percolation model with a huge variety of
applications in many fields were introduced. An important development first established by Fortuin and Kasteleyn, is the connection between bond percolation and a lattice statistical model. This consideration has led to a formulation of the percolation problem as a limiting
case of the general Potts model,
which was extremely useful, for many of the techniques readily available in statistical mechanics has been applied to percolation \cite{Wu1978JSPPercolation}.  
Many variations were motivated by consideration of spatial correlations, 
anisotropy, nonlocality, explosivity  etc. 'Explosivity' in particular, deals 
with the search for those perturbations of cluster-merging rules which change 
the order of percolation transition from the second order (continuous) to the 
first order (discontinuous). This is known as \textit{explosive percolation} in 
which a macroscopic connected
component emerges in a number of steps that is much smaller than the system size. This has recently become a subject of enormous interest \cite{Bastas2014explosive}, including openings toward other subjects such as jamming in the Internet \cite{Martino2009PRE}, synchronization phenomena \cite{Leyva2013Explosive, Moreno2011PRL}
and analysis of real-world networks \cite{Pan2011PRE}. Let us first start by introducing the family of explosive percolation models and then turn our attention to some other variants and modifications. The number of papers appeared on explosive percolation was itself explosive and we cannot cover all subjects and ideas of course, we rather try to outline some original and selective topics. 

\subsection{Explosive percolation}

\subsubsection{Achlioptas process}\label{Achlioptas}

Having introduced in subsection~\ref{randomgrapghs}, the classic Erd\H{o}s-R\'{e}nyi random graph (or briefly ER$_n$ model), is composed of $n$ isolated vertices whose each pair of vertices is chosen uniformly at random  in each step, and connected by an edge \small\{e$_1$\}. At any given moment, a cluster is defined as the set of vertices each of which
can be reached from any other vertex in it by traversing edges. If $tn$ denotes for the number of added edges at time $t$, it is known that the fraction of vertices in the largest cluster undergoes a continuous phase transition at the critical time $t_c=1/2$.

At a Fields Institute workshop in 2000, Dimitris Achlioptas suggested a class of variants of the classical ER$_n$ model where a nonrandom selection rule, the so-called Achlioptas process, is additionally imposed which tends to the delay (or acceleration) in the formation of a large percolating cluster. This has then received much attention in recent years \cite{Bohman2001Random, Spencer2007Comb, Beveridge2007Proc, Krivelevich2008Hamiltonicity}. Concretely, consider a model that,
like ER$_n$, starts with $n$ isolated vertices and add edges one by one. The difference is that to add a single edge, first two random edges \small\{e$_1$,e$_2$\small\} are chosen, rather than one, each edge is chosen exactly as in ER$_n$ and independently of the other. Of these, only one should be selected according to the selection rule, and then inserted in the graph. The other edge is discarded. Clearly,
if one always resorts to randomness for selecting
between the two edges, the original ER$_n$ model is recovered.

Achlioptas originally asked if it is possible to shift the critical point of this phase transition by following an appropriate selection rule. One rule that can naturally be imagined is the product rule: Of the given potential edges, pick the one which minimizes the product of the sizes of the components containing the four end points of \small\{e$_1$,e$_2$\small\}. This rule was suggested in \cite{Bollobas1984Random} as the most likely to delay the critical point. Another rule is the sum rule, where the size of the new component formed is minimized.\\ A selection rule can be classified as a bounded-size or an unbounded-size rule. In a bounded-size selection rule, decisions depend only on the sizes of the components and, moreover, all sizes greater than some (rule-specific) constant $K$ are treated identically. For example $K=1$ is the Bohman-Frieze (BF) rule, where $e_1$ is chosen if it joins two isolated vertices, and $e_2$ otherwise. It was rigorously proven in \cite{Bohman2001Random} for a much simpler rule, that such rules are capable to shift the threshold. Moreover, the percolation transition is strongly conjectured to be continuous for all bounded-size rules \cite{Spencer2007Comb}. For unbounded-size rules in contrast, extensive simulations \cite{Achlioptas2009Explosive} strongly suggested that, the product rule in particular,
shows much more interesting behavior than just shifting the critical point,
it exhibits a discontinuous percolation transition. The numerical evidence showed that the fraction of vertices
in the largest cluster jumps from being a vanishing
fraction of all vertices to a majority of them instantaneously, i.e., the largest cluster grows from size at most $\sqrt{n}$ to size at least $n/2$ in at most $2n^{2/3}$ steps, that is, at the phase transition a constant fraction of the vertices is accumulated into a single giant cluster within a sublinear number
of steps.

However, this conjecture has been rigorously disapproved by Riordan and Warnke 
\cite{Riordan2011Science, Riordan2012AAP} in 2011, by showing that it cannot 
be a discontinuous transition, but a continuous one. In fact, their 
argument shows continuous
phase transitions for an even larger class of processes \cite{Riordan2012AAP} called $l-$vertex rules (every Achlioptas
process is a $4-$vertex rule).
Their results state that continuity of the phase transition is
such a robust feature of the basic model that it
survives under a wide range of deformations and thus all Achlioptas processes have a continuous
phase transition. This however neither means that the connectivity transition at $p_c$ is characterized by a usual power-law divergence of the order parameter \cite{Stanley1971Book, Sornette2006Book, daCosta2010PRL} nor that it cannot be followed by multiple discontinuous transitions. In particular, Nagler et al.\  have proven that for certain $l-$vertex rules the continuous phase transition can have the shape of an incomplete devil's staircase with discontinuous steps in arbitrary vicinity of $p_c$ \cite{nagler2012PRX}. \\Therefore, heuristics
or extrapolations from simulations suggesting explosive percolation in mean-field models seem to be wrong in the scaling limit. Nevertheless, a discontinuous transition is possible when one departs far enough
from the ER$_n$ model.

\subsubsection{Half-restricted process}

Half-restricted process \cite{Panagiotou2011Explosive} is a variant of the Erd\H{o}s-R\'{e}nyi process which exhibits a discontinuous phase transition. Intuitively, a discontinuous transition can only occur if the growth of large clusters is suppressed and clusters with medium sizes become abundant. 
After a number of steps, such medium-sized clusters merge and a giant cluster emerges drastically.
As we saw in the last subsection~\ref{Achlioptas}, the idea to do this in an Achlioptas
process was to select an edge that connects smaller
components, which did not lead to an explosive percolation. A different approach is however considered in the half-restricted process: In each step, two vertices are connected by an edge, but one of them is restricted to be within the smaller components.
\begin{figure}[t]
\centerline{
\includegraphics[width=1.06\textwidth]{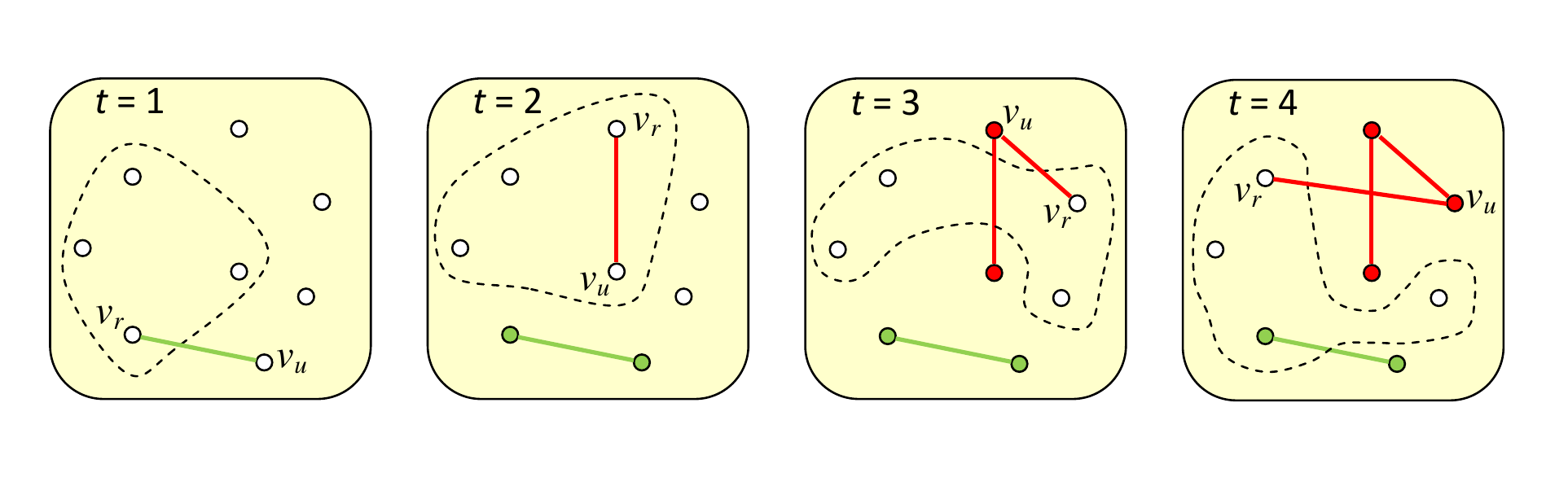}
}
 \caption{Illustration of the first four steps of time evolution of a graph $G_n$ with $n=8$ number of vertices, according to the half-restricted process. The vertices surrounded by the dashed line belong to the restricted vertex set $R_{f}(G)$ with $f=0.5$.} 
\label{fig8}
\end{figure}

Consider a graph $G_n$ including $n$ labeled vertices $v_1, v_2, \dots, v_n$ sorted ascending in the size of the clusters that they reside in. Vertices with the same cluster size are sorted lexicographically. 
For a given value of $0<f\le1$, a restricted vertex set $R_f(G)$ is defined which is composed of the $\lfloor fn\rfloor$ number of vertices in smallest components ($\lfloor fn\rfloor$ means the floor of $fn$, i.e., the largest integer less than or equal to $fn$). At each time step $t\ge 1$, one edge is added between two chosen vertices $v_r$ and $v_u$. Starting with an empty graph at $t=1$, at every step $t$, first the restricted vertex set $R_f(G)$ is recognized, and one restricted vertex $v_r$, which belongs to $R_f(G)$, is uniformly chosen at random. Then independently, one unrestricted vertex $v_u$ is chosen uniformly at random from the whole vertex set $G_n$. An edge is then added between two vertices $v_r$ and $v_u$ if is not already present. Figure~\ref{fig8} illustrates the first four steps of time evolution of an $G_n$ graph with $n=8$ and $f=0.5$, according to the half-restricted process.

It has been shown in \cite{Panagiotou2011Explosive} that, unlike the Achlioptas process, the half-restricted process
exhibits explosive percolation with a discontinuous phase transition for any value of $f<1$. Although the evolution of the largest cluster over the first $n$ edges in the half-restricted process may behave very similar to that of the Achlioptas process \cite{Panagiotou2011Explosive}, they are fundamentally different in the nature of phase transition as well as  mathematical structure.

\subsubsection{Spanning cluster-avoiding process}

For the continuous phase transitions, the Erd\H{o}s-R\'{e}nyi process on random graphs accounts for the mean-field limit of the standard percolation on a Euclidean lattice. Explosive percolation models in Euclidean space have also been extensively studied \cite{ziff2009PRL, Ziff2010PRE}, and the numerical
results suggest discontinuous transitions \cite{Choi2011PRE}. However, due to the lack of analytic arguments, the
order of explosive transition in Euclidean space is still elusive. This would therefore be of interest to clarify the order of the explosive
transition in Euclidean space and on random graphs in a unified manner. To this aim, a model called the spanning cluster-avoiding (SCA) model was introduced \cite{ZiffScience2013, Cho2013Science}. The spanning cluster in Euclidean space actually plays the same role as the giant cluster (or component) in random graph models. The SCA model starts by considering a finite hypercubic lattice $\mathbb{Z}^d$ in $d$ dimensions of size $N$ and unoccupied bonds. Inspired by the best-of-$m$ model \cite{Hermann2011Tricritical}, at each time step $t$, number of
$m$ unoccupied bonds are chosen randomly and classified into two types: bridge and nonbridge bonds. Bridge bonds
are those that upon occupation a spanning cluster is formed. SCA model avoids bridge bonds to be occupied, and thus one of the nonbridge bonds is randomly selected and occupied. If the
$m$ potential bonds are all bridge bonds, then one of them is randomly chosen and occupied. Once a spanning cluster is formed, restrictions are no longer imposed on the occupation of
bonds. This procedure continues until all bonds
are occupied at $t=1$.

Extensive numerical simulations and theoretical results \cite{Cho2013Science} have shown that the explosive transition in SCA model in the thermodynamic limit, can be either discontinuous
or continuous for $d<d_c=6$
depending on the number of potential bonds $m$.
In other words, there is a tricritical value $m_c(d)=d/(d-d_{BB})$ for $d>d_{BB}$, where $d_{BB}$ denotes for the fractal dimension of the set of bridge bonds \cite{Schrenk2012Fracturing}, such that if $m<m_c$, the transition is continuous at a finite threshold $t_c$, and discontinuous, in the thermodynamic limit, for $m>m_c$ at the trivial percolation threshold at $t=1$, when all bonds of the system are occupied. 
The formula for $m_c(d)$ is valid only for $d<d_c=6$ at which $m_c(d_c=6)=\infty$.  \\For $d\ge d_c$, i.e.,
in the mean-field limit, the transition is shown to be continuous for any finite and fixed value of $m$. However if $m$ varies with the
system size $N$, a discontinuous transition can also take place for $d\ge d_c$. More precisely, there exists a characteristic value $m_c\sim \ln N$ such that when $m$ increases with $N$
slower than $m_c$, the transition is continuous in a finite critical time $t_c$, and when $m$ increases with $N$ faster than $m_c$, the transition is discontinuous at the trivial percolation threshold at $t=1$. If $m$ increases with $N$ as $m_c\sim\ln N$, then a discontinuous transition occurs at finite $t_{cm}$, which is neither $t_c$ nor unity. Some necessary conditions for a non-trivial
and a trivial discontinuous percolation transition have been recently proposed \cite{Cho2014arXivOrigin}.

The idea to obtain a discontinuous percolation transition by controlling the largest cluster alone was actually addressed before 
\cite{Hermann2010Explosive}, in which an acceptance method was used to systematically
suppress the formation of a largest cluster on the lattice. There has been also shown that the cluster perimeters are fractal at the percolation threshold with a fractal dimension of $1.23\pm0.03$, statistically indistinguishable from that of watersheds.

\subsection{Non-self-averaging percolation}
\begin{figure}[t]
 \centerline{
 \includegraphics[width=0.6\textwidth]{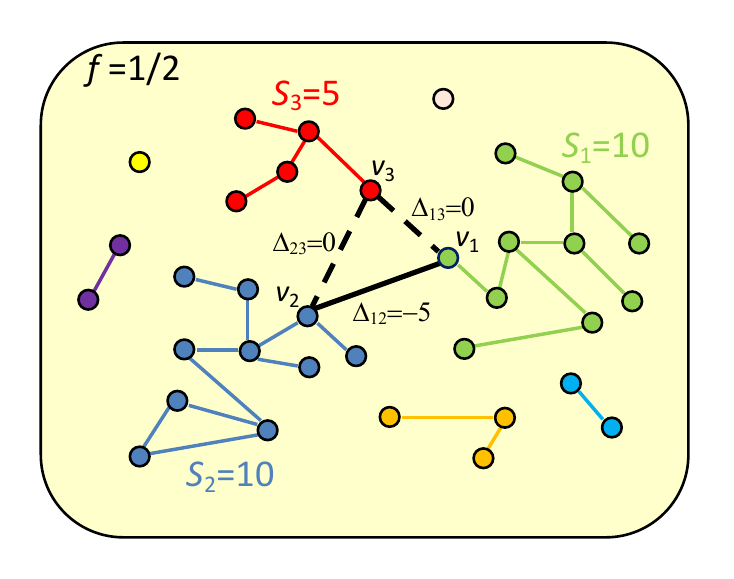}
 }
  \caption{Illustration of the cluster merging process according to the fractional percolation with target fraction $f=1/2$. The link between vertices $v_1$ and $v_2$ is established because $\Delta_{12}=-5$ is minimal. } 
 \label{fig9}
 \end{figure}

 Fractional percolation, as a non-self-averaging percolation, was introduced 
 \cite{Habib2013Nature} to describe some of the main features of crackling 
 noise as a characteristic feature of many systems when pushed slowly, e.g., 
 the crumpling of paper \cite{Houle1996PRE}, earthquakes 
 \cite{Gutenberg1954Book} and the magnetization of slowly magnetized magnets. 
 In the fractional percolation the relative size of the largest component 
 $s_{max}$, as the order parameter, exhibits many randomly distributed jumps 
 after a critical threshold $p_c$ whose discontinuities survive even in the 
 thermodynamic limit. A fractional growth rule is used in this model which 
 induces a certain type of size homophily among clusters: Connection of two 
 clusters with a similar size is in priority, while the size of the larger 
 cluster has already been rescaled by a target fraction factor $0<f\le 1$. The 
 model starts by considering a network of $n$ isolated vertices with no edges. 
 At each step, three different vertices $v_1, v_2, v_3$ are chosen uniformly at 
 random, no matter if they are in the same cluster. Let $S_1\ge S_2\ge S_3$ 
 denote the sizes of the (not necessarily distinct) clusters that they are 
 contained in. Then an edge is added between the two vertices $v_i$ and $v_j$ 
 for which $\Delta_{ij}:=fS_i-S_j, 1\le i<j\le 3$ is minimal. If there are 
 multiple options, one is randomly chosen. Moreover when only two clusters are 
 left in the system they will be connected---see Fig.~\ref{fig9} for further 
 illustration. 
 
 It has been shown that after the first transition, for $p>p_c$ and $n\rightarrow\infty$, the size of the largest component either stays constant or increases fractionally by at least a factor of $r=f/(1+f)$. As the
 first transition is point-continuous \cite{Riordan2011Science, Aharony1996PRLAbsence}, the process necessarily undergoes infinitely many discontinuous transitions arbitrarily close to the first dynamical transition point $p_c$, and the order parameter  increases stepwise. The height $\delta_k$ of the $k$th step down the staircase has the form $\delta_k\sim (1+r)^{-k}$, giving rise to the scaling distribution $d_s$ of the jump size as $d_s\sim s^{-1}$. Both the size of the jumps, and the transition points are stochastic even in the thermodynamic limit.\\The process is therefore shown to be non-self-averaging in the sense that the relative variance of the size of the largest component given by $\langle s_{max}^2\rangle/\langle s_{max}\rangle^2-1$, does not vanish for $p\ge p_c$ as $n\rightarrow\infty$. This
 is in contrast to classical models (e.g., the ER$_n$ model) for which the order parameter converges to a nonrandom function in the thermodynamic limit. \\The characteristics of fractional percolation are robust against
an arbitrary variation of the target fraction factor $f$. Assuming a time-dependence form of $f(k)=\alpha/k$ where $0<\alpha\le 1$, one finds that the jump size distribution function decays faster than $\sim s^{-1}$ characterized by the power-law fluctuations $d_s\sim s^{-\beta}$ with $\beta=(1+\alpha)/\alpha$.

For other models that exhibit non-self-averaging in percolation we refer to the literature \cite{riordan2012, chen2013a, chen2013b, naglerPRL2014}.

\subsection{Correlated percolation}\label{corrPerc}

In the percolation models studied so far, all random occupations have been considered to be independent of each other with no spatial correlations. But this is not always the case when, for instance, percolation theory is applied to study transport and geometric properties of disordered systems \cite{Du1996AIchE, Coniglio1979PRL, Makse1995Nature, Makse1998PRE, Makse2000PRE, Makse2002PRETraveling, Moreira2003PRE}, since the presence of disorder usually introduces spatial correlations in the model. For sufficiently short-range correlations, the properties of the model will be the same as those of uncorrelated percolation. By increasing the range of correlations however, they may have a relevant contribution leading to new
fixed points in the renormalization group study of the model. These can be quantified in terms of how the correlation function $g(r)$ falls off at large distances $r$: When the correlations are short-range with a fall-off faster than $r^{-d}$, then according to the Harris criterion \cite{HarrisCriterion1974}, they are relevant if $d\nu-2<0$, where $\nu$ is the correlation length exponent for the uncorrelated percolation model. Since for the percolation model we always have the hyperscaling relation $d\nu-2=-\alpha>0$, so short-range correlations do not change
the critical behavior \cite{Weinrib1983PRB}. For long-range corralations of the power-law form $g(r)\sim r^{-2H}$ with $2H<d$ instead, the extended Harris criterion \cite{Weinrib1984PRB} states that the correlations are relevant if $H\nu -1<0$. In this case, the new correlation length exponent is given by the scaling relation $\nu_H=1/H$. This relation has been then verified numerically in \cite{SaberiAPL, Makse2000PRE, Marinov2006PRE, Prakash1992PRA, Abete2004PRL}. Thus the critical exponents in a long-range \textit{correlated percolation} can change depending on how the correlations decay with the spatial distance. 

Such a power-law decay of the spatial correlations is a typical characteristic feature of the height profile $\{h(\textbf{x})\}$ in random grown surfaces,
where $h(\textbf{x})$ is the height at the lattice site at position $\textbf{x}$. This indeed provides a convenient way to tackle with the correlated percolation on a lattice. For self-affine surfaces for which $g(r)\sim (1-r^{2H})$, it is shown in \cite{Schmittbuhl1993JPA} that even in the thermodynamic limit, the percolation transition is only critical for $H=0$. Percolation on these surfaces is actually governed by the largest wavelength of the height distribution, and thus the self-averaging breaks down. For long-range correlated surfaces where $g(r)\sim r^{-2H}$, in contrast, the transition is critical and the self-averaging is recovered. Depending on the value of $H$, the correlation-length exponent is given as follows
\begin{equation}
\nu_H=
\left\{
\begin{array}{c c c}
1/H  &  \mbox{if} &  0<H<1/\nu,
\\
\nu &  \mbox{if} &\qquad  H\ge 1/\nu, 
\end{array} \right.
\end{equation}
where $\nu=4/3$ for a Euclidean lattice in 2D. It is thus natural to expect that other critical exponents also depend on the value of $H$. Such a dependence is numerically verified for the fractal dimensions of the largest cluster $d_{fH}^c$, its perimeter $d_{fH}^P$ and external perimeter $d_{fH}^{EP}$, shortest path, backbone, and red sites \cite{PhDthesisScrenk}. It has also been shown that, within the numerical accuracy, the hyperscaling relation $d=(\gamma_H+2\beta_H)/\nu_H=\gamma_H/\nu_H+2(d-d_{fH}^c)$ is fulfilled by the exponents. The duality relation is numerically shown to be valid $(d_{fH}^{EP}-1)(d_{fH}^{P}-1)=1/4$, though the theoretical verification of these observations is still lacking.

\subsection{Bootstrap percolation}

The bootstrap percolation problem \cite{Alder1991PA, Alder2003BJP} and its obvious variants deal with the dynamics of a system composed of highly coupled elements, each of which has a state that depends on those of its close
neighbors. It has played a canonical role in description of a growing list of complex phenomena including crack propagation \cite{Alder1988JPA},  neuronal activity \cite{Eckmann2007PhysRep, Soriano2008NAS, Goltsev2010PRE}, and magnetic systems \cite{Sabhapandit2002PRL} among others.
\\The standard bootstrap percolation process on a lattice as the spread of activation or infection is defined according to the following rule with a given fixed parameter $k\ge2$: Initially, each of the sites is randomly infected (or activated) with probability $p$ and uninfected (inactivated) with probability $1-p$, independently of the state of the other sites. Every infected site remains infected forever, while each uninfected one which has at least $k$ infected neighbors becomes infected and remains so forever. This procedure is continued until the system reaches the stable configuration which does not change anymore i.e., when no uninfected site has $k$ or more infected neighbors. The main question which arises is concerning the percolation of the infected cluster i.e., if there emerges a giant spanning cluster of infected sites of size $\mathcal{O}(n)$ by the end of the process.

Bootstrap percolation has been extensively studied on 2D and 3D lattices \cite{Holroyd2003TRF, Holroyd2006EJP, Balogh2006PTR, Cref1999AnnProb} (and references therein), including the proof of the existence of a sharp metastability threshold in 2D \cite{Holroyd2003TRF} which has then been generalized to arbitrary $d$ dimensional lattices \cite{Holroyd2006EJP, Balogh2006PTR}. In particular, Schonmann \cite{Schonmann1992AnnProb} proved that on the infinite lattice $\mathbb{Z}^d$, the percolation threshold $p_c(\mathbb{Z}^d, k)=0$ if $k\le d$,  and $p_c(\mathbb{Z}^d, k)=1$ otherwise. The finite size behavior (also known as metastability) was studied in \cite{Cref1999AnnProb, Aizenman1988Metastability, Cerf2002SPA}, and the threshold  function was determined up to a constant factor, for all $2\le k\le d$, by Cerf and Manzo \cite{Cerf2002SPA}. The first sharp threshold was determined by Holroyd \cite{Holroyd2003TRF}, for $k=2$ on a finite 2D lattice $\mathbb{Z}^2$ of linear size $L$, who proved that\begin{equation}\label{2Dpc}
p_c(L, d=2, k=2)\simeq\dfrac{\pi^2}{18\ln L},\qquad \text{as}\quad L\rightarrow\infty.
\end{equation}
This has been recently \cite{Balogh2012TAMS} generalized to the finite lattice $\mathbb{Z}^d$ of linear size $L$ as follows \begin{equation}\label{dpc}
p_c(L, d, k)\simeq\bigg(\dfrac{\lambda(d,k)}{\ln_{(k-1)}L}\bigg)^{d-k+1},\quad \text{for}\quad 2\le k\le d, \quad \text{as}\quad L\rightarrow\infty,
\end{equation}
where $\ln_{(k)}$ denotes for an $k-$times iterated logarithm, $\ln_{(k)}L=\ln\big(\ln_{(k-1)}L\big)$.
Although the function $\lambda(d,k)$ is not exactly known, but it can be shown that it always has a finite value with the following properties \cite{Balogh2012TAMS}: $\lambda(2,2)=\pi^2/18$, $\lambda(d,2)\simeq (d-1)/2$, and $\lambda(d,d)\simeq\pi^2/6d$ as $d\rightarrow\infty$. Clearly the result (\ref{2Dpc}) is a special case of (\ref{dpc}) for $d=2$ and $k=2$.
Surprisingly, there exist some predictions for the asymptotic threshold in 2D based on numerical simulations which differ greatly from the rigorous result (\ref{2Dpc}). For example, based on simulations in \cite{Alder1989JPA}, the estimate
$p_c(L,d=2,k=2)\ln L = 0.245\pm0.015$ is reported whereas the rigorous prediction is $\pi^2/18= 0.548311\dotsb$. This apparent discrepancy between theory and experiment has been rigorously addressed in \cite{Gravner2008AAP} to be due to a very slow convergence in the asymptotic limit when $L\rightarrow\infty$.

Bootstrap percolation has also been studied on the random regular graph \cite{Balogh2007RSA, Fontes2008JSP}, and
infinite trees \cite{Balogh2006CPC} as well. In the context of real-world networks and in particular in social networks, a bootstrap
percolation process can be imagined as a toy model for the spread of ideas or new trends within a set of individuals which form a network. In this sense, the bootstrap percolation has been recently \cite{Amini2014JSP} studied on the power-law random graphs of $n$ vertices in which is shown that a giant cluster of size $\mathcal{O}(n)$ emerges, with high probability, above a certain critical sublinear number of the initially infected nodes. This behavior is in sharp contrast with that observed \cite{Balogh2007RSA, Janson2012AAP, Baxter2010PRE} on an Erd\H{o}s-R\'{e}nyi random graph in which there will be no evolution when the number of initially infected vertices is sublinear.

The final remark is regarding the close relation between the bootstrap percolation and another well-known problem in graph theory, that of the $k$-core of random graphs \cite{Dorogovtsev2006, Bollobas1984Proc, Pittel1996JCTSB}.  The $k$-core of a graph is the maximal subgraph for which all nodes have at least $k$ neighbors within the $k$-core. However there is a difference between the stationary state of the bootstrap percolation and the $k$-core \cite{Baxter2010PRE}. Bootstrap percolation is an infection process, which starts from a subset of source nodes and spreads over a network according to the infection rules described earlier. The $k$-core of the network can be found as an asymptotic structure obtained by a subsequent pruning of nodes which have less than $k$ neighbors.

\subsection{Directed percolation}

Considering a porous rock as a random medium in which neighboring pores are connected by small channels of varying permeability, an important problem in geology would be how deep the water can penetrate into it. Clearly the ordinary percolation model is not applicable to describe such phenomena since the gravity has weighted a specific direction in space i.e., the water propagation is not isotropic but \textit{directed}.
 \begin{figure}[t]
 \centerline{
 \includegraphics[width=0.8\textwidth]{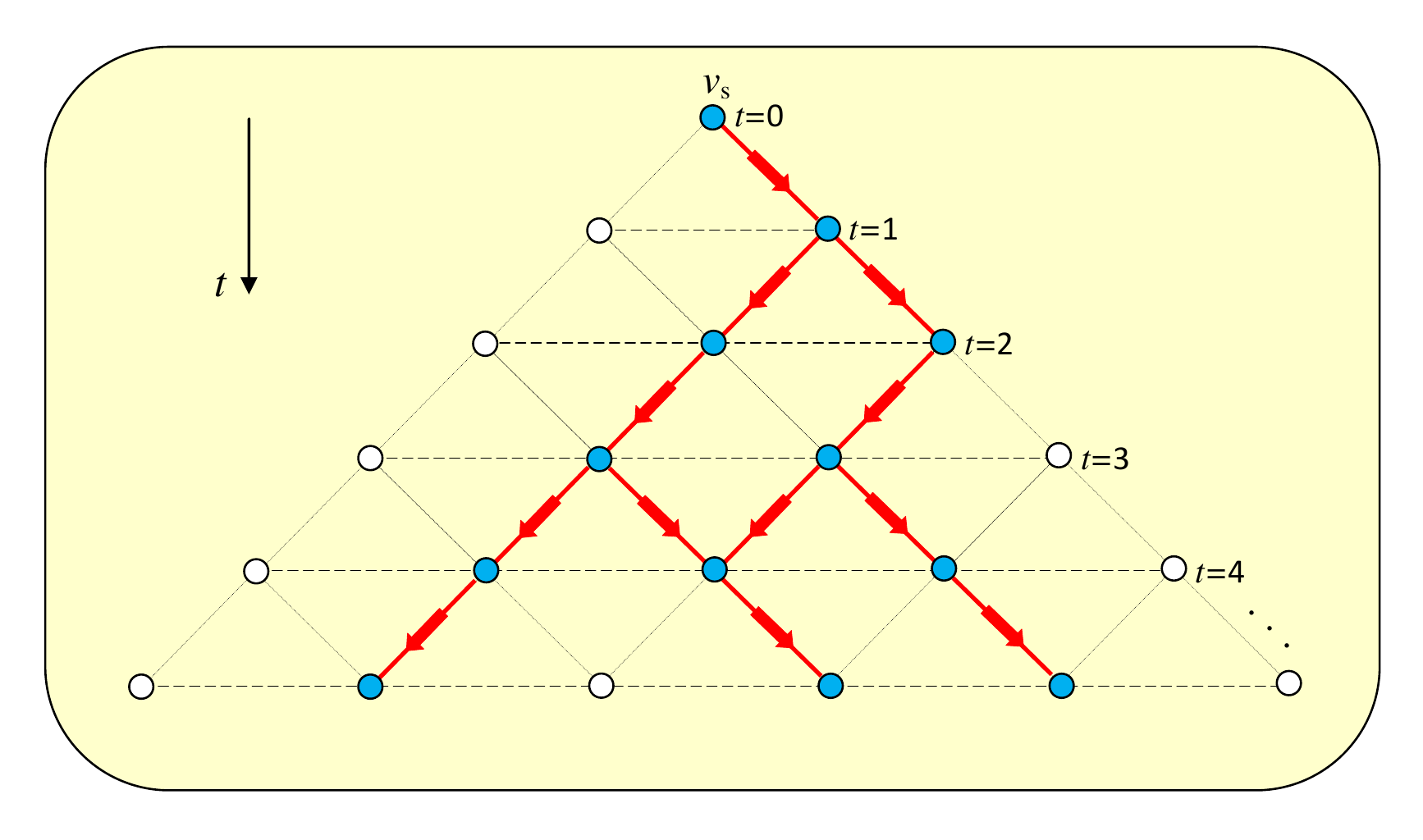}
 }
  \caption{Directed bond percolation on a tilted square lattice started from a source active vertex $v_s$ at $t=0$. Each bond (and corresponding destination site) is activated with probability $p$, shown by solid lines with arrows (filled circles). The cluster of active sites connected by a directed path to $v_s$ is indicated in blue (dark).
  The vertical direction corresponds to time, and the dashed
  lines identify the sets of available vertices at time $t$. } 
 \label{fig10}
 \end{figure}
Directed percolation, introduced in \cite{Broadbent1957}, is an
anisotropic variant of standard isotropic percolation which introduces a specific direction in space. It accounts for one of the most prominent universality classes of \textit{nonequilibrium} phase transitions, playing a similar role as the Ising universality class in \textit{equilibrium} statistical mechanics \cite{Hinrichsen2000AdvPhys}. Directed percolation models exhibit a continuous phase transition with a fascinating property of robustness with respect to the microscopic dynamic rules. It turned out to describe a wide range of spreading models e.g., contact process \cite{Liggett1985Book, Dickman1988PLA},
epidemic spreading without immunization \cite{Grassberger1982MB}, and forest fire models \cite{Nahmias1989RPA, Albano1995PA, Albano1994JPA}. For this model, the percolation can only occur along a given spatial direction. Regarding this direction as a temporal degree of freedom, directed percolation can then be viewed as a dynamical process in $d+1$ dimensions. In this sense percolation on a dynamic network can be mapped onto the problem of directed percolation in infinite dimensions \cite{Parshani2010EPL}.

Starting from a source active (occupied or wet) vertex $v_s$ at $t=0$ on a 
tilted hypercubic $\mathbb{Z}^d$ lattice, directed bond percolation, as a 
dynamical process, can be interpreted as follows. As illustrated in 
Fig.~\ref{fig10}, at the next time step, each of the downward bonds emanating 
from $v_s$ is randomly occupied by an arrow with probability $p$ which 
corresponds to the destination site to become active. This procedure is 
continued row by row until the system reaches an \textit{absorbing state}, 
i.e., a configuration that the model can reach but it cannot escape from there. 
There exists a critical threshold $p_c$, that for $p<p_c$ the average number of
active sites grows for a short time and then decays exponentially. For $p>p_c$ there is a
finite probability that the number of active sites diverges as $t\rightarrow\infty$. In this case activity spreads within a so-called spreading cone. At $p=p_c$ which separates a non-fluctuating absorbing state from a fluctuating active phase, a critical cluster is
generated from a single source whose scaling properties are characterized by a number of critical exponents.
Note that, in directed percolation each source vertex generates an individual cluster, so the lattice in this case cannot be decomposed into disjoint clusters. 
\\Therefore, depending on the value of $p$, activity may either spread over the entire system or die out after some time. The latter case i.e., the absorbing state is a completely inactive state which can only be reached by system but not be left. Thus detailed balance is no longer obeyed and that is why the process is nonequilibrium. \\
Despite of its very simple rules and robustness, its critical behavior is
highly nontrivial and exact computation of the critical exponents seems impossible. Even in $1+1$ dimensions, no analytical solution is known, suggesting that is a non-integrable process. 
This may be related to the fact that directed percolation, unlike the ordinary percolation, due to the lack of symmetry between space and time, is not conformally invariant.

Although it is not an equilibrium model, the critical behavior of the directed percolation shares a very similar picture as in the ordinary percolation. A phenomenological scaling theory can be applied to describe its criticality. Considering the density $\rho(t)$ of active sites as an order parameter of a spreading process, observations justify that in the active phase $\rho(t)$ decays and eventually saturates at some stationary value $\rho_s(t)$. Near the critical point, the stationary density is then turned out to satisfy a power-law relation $\rho_s\sim(p-p_c)^\beta$, where $\beta$ is a universal critical exponent that only depends on dimensionality. The other important quantity is the correlation length whose definition needs a special care, since in this case, time is an additional dimension which should be distinguished from the spatial dimensions. Let us denote the temporal and spatial correlation lengths by the indices $\xi_\parallel$ and $\xi_\perp$, respectively, which are independent of each other. Then close to the transition, these length scales are expected to diverge as $\xi_\parallel\sim|p-p_c|^{\nu_\parallel}$ and $\xi_\perp\sim|p-p_c|^{\nu_\perp}$ with  generally different critical exponents $\nu_\parallel$ and $\nu_\perp$. The two correlation lengths are related in the scaling regime by $\xi_\parallel\sim\xi_\perp^z$ where $z=\nu_\parallel/\nu_\perp$ denotes for the so-called dynamic exponent. In many models, the universality class is given by the three exponents $\beta$, $\nu_\parallel$ and $\nu_\perp$. \\For $d\ge d_c=4$, the values of the critical exponents are believed to be given by the mean-field theory as $\beta=1$, $\nu_\parallel=1$ and $\nu_\perp=1/2$ \cite{Hinrichsen2000AdvPhys}. For $d<d_c$ however, there are no
exact results neither for critical exponents nor thresholds. However, a very precise estimates of critical exponents and thresholds on several lattices can be found in \cite{Jensen1996JPA, Jensen1999JPA} in $1+1$ dimensions, and in a recent work \cite{Wang2013PRE} (and references therein) in higher dimensions. 

Let us conclude this subsection by stating the following conjecture first made by Janssen \cite{Janssen1981ZPB} and Grassberger \cite{Grassberger1982ZPB} inspired by the variety and robustness of directed percolation models. The statement is that any model which fulfills the following conditions should belong to the directed percolation universality class: i) The model exhibits a continuous phase transition into a unique absorbing state, ii) The transition is characterized by a positive one-component order parameter, iii) The dynamic rules involve only short-range interactions, and iv) The system has no special attributes e.g., additional symmetries or quenched randomness. Although this conjecture is not yet rigorously proven, it is highly supported by numerical evidence.

\section{Percolation in two dimensions}
\label{chap4}

At the critical point, the correlation length diverges and the system becomes scale invariant and thus scaling hypothesis applies. However the scaling hypothesis alone cannot determine the critical exponents though it may give some relations among them. Therefore one may call for some possible stronger symmetries e.g., conformal invariance, to determine the exponents. Of course scale invariance does not imply conformal invariance at least at the level of the superficial mathematical definition. How can thus conformal invariance to be enhanced from scale invariance? In two dimensions, one can rigorously show that scale invariance is enhanced to conformal invariance under the following assumptions \cite{Polchinski1988NPB, Zamolodchikov1986JETP, Nakayama2014arXiv}: i) unitarity, ii) Poincar\'e invariance (causality), iii) discrete spectrum in scaling dimension, iv) existence of scale current, and v) unbroken scale invariance. Fortunately most interesting classes of $1+1$ quantum field theories, as the scaling limit of various 2D lattice models in statistical mechanics, satisfy these assumptions. In two dimensions however the conformal symmetry is extremely powerful since conformal transformations correspond to analytic functions to be used in statistical mechanics to characterize universality classes. Under conformal transformations the lengths are rescaled non-uniformly while the angles between vectors are left unchanged.

Many exact results have then been obtained for percolation model in two dimensions using methods of conformal field theory (CFT). Among them, Cardy's conjectured formula \cite{Cardy1992JPALCrossing, Cardy2002arXivCross} for the crossing probability is one of the famous ones. For a percolation model defined in a unit disc $|z|<1$,  the probability that there exists at least one cluster which contains at least one point from each of two disjoint intervals on the boundary of the disc whose ends are assigned by ($z_1,z_2$) and ($z_1,z_2$) respectively, has the following explicit form as a function of the cross-ratio $\eta$ \begin{equation}\label{CrossingProb}
P_s\big((z_1,z_2),(z_3,z_4);\eta\big)=\dfrac{\Gamma(\frac{2}{3})}{\Gamma(\frac{4}{3})\Gamma(\frac{1}{3})}\eta^{\frac{1}{3}}\leftidx{_2}{F_{1}}(\frac{1}{3},\frac{2}{3};\frac{4}{3};\eta), \qquad \eta\equiv\frac{(z_1-z_2)(z_3-z_4)}{(z_1-z_3)(z_2-z_4)},
\end{equation}
where $\leftidx{_2}{F_{1}}$ is the hypergeometric function. $P_s(\eta)$ is invariant under transformations of the unit disc which are conformal
in its interior (but not necessarily on its boundary). Of course, this probability is interesting only at the critical point. For $p<p_c$, since all clusters are finite, in the scaling limit we will have $P_s=0$, while for $p>p_c$ the infinite cluster always spans, so the limit is $1$. This result is valid as long as there are only short-range correlations in the probability measure, independent of whether the microscopic model is formulated as bond, site, or any other type of percolation. \\Moreover it has been shown that the whole probability
distribution of the total number
of such distinct crossing clusters is conformally invariant in the scaling limit, depending only on $\eta$. In particular the mean number $\mathcal{N}_{c}$
of such crossing clusters is \cite{Cardy2001Lecturenote} \begin{equation}
\mathcal{N}_{c}(\eta)=\frac{1}{2}-\frac{\sqrt{3}}{4\pi}\bigg[\ln(1-\eta)+2\sum_{m=1}^{\infty}\dfrac{\Gamma(\dfrac{1}{3}+m)\Gamma(\frac{2}{3})}{\Gamma(\dfrac{2}{3}+m)\Gamma(\frac{1}{3})}\dfrac{(1-\eta)^m}{m}\bigg].
\end{equation}

The other exact result obtained by using conformal field theory and Coulomb gas methods, is for the number $\mathcal{N}(A)$ of percolation clusters of enclosed area greater than or equal to $A$ at the critical point in two dimensions \cite{CardyZiff2003JSP}. It has been shown to behave as the following power-law \begin{equation}
\mathcal{N}(A)\sim \dfrac{C}{A},
\end{equation}
with a proportionality constant $C=1/8\sqrt{3}\pi$ that is universal.

A rigorous proof of Cardy’s
formula for site percolation on the triangular lattice was discovered \cite{Smirnov2001Proof} by Smirnov using advantages of the theory of stochastic L\"{o}wner evolution (SLE).

\subsection{Stochastic L\"{o}wner evolution}

In probability theory, the stochastic L\"{o}wner evolution (SLE$_\kappa$) with parameter $\kappa$, introduced by Schramm \cite{Schramm2001SLE} in 2000, is a family of random planar curves that have been proven to be the scaling limit of interfaces in a variety of 2D critical lattice models in statistical mechanics. Here we give a brief description of the so-called chordal SLE in its standard setup, and refer to \cite{Cardy2005SLE, Bernard2006SLE, Nienhuis2004SLE} for more detailed information about it, and other variants of SLE.

Consider a curve $\gamma(t)$ that emanates at $t=0$ from origin on the boundary of the upper half-plane $\mathbb{H}$, and goes to infinity as $t\rightarrow\infty$. The curve might intersect but should not cross itself during the evolution. Let us define the hull $K_t$ as the union of the curve and the set of points got trapped by the curve up to time $t$, which are not reachable from infinity without crossing the curve. According to the Riemann mapping theorem, there should be an analytic function $g_t(z)$ which maps $\mathbb{H}\backslash K_t$ into the $\mathbb{H}$ itself ($\mathbb{H}\backslash K_t$ is a simply connected domain that contains all those points of $\mathbb{H}$ that are not in $K_t$). The map $g_t(z)$ can be uniquely determined by imposing the following hydrodynamic normalization at infinity,\begin{equation}
g_t(z)=z+\dfrac{2t}{z}+\dotsb,\qquad \text{as}\quad z\longrightarrow\infty,
\end{equation}
where the coefficient $2t$ is due to a conventional parameterization of $\gamma$. It can be then shown that the evolution of the tip of the curve, can be given by the L\"{o}wner differential equation
\begin{equation}
\partial_t g_t(z)=\dfrac{2}{g_t(z)-\zeta_t}, \qquad g_0(z)=z, \quad z\in\mathbb{H},
\end{equation}
where $\zeta_t$ is a continuous function but not necessarily differentiable. In order to have conformally invariant random curves which behave geometrically as they should to encode the statistics of critical interfaces, Schramm argues that \cite{Schramm2001SLE} they should have two properties: Markov property, and stationarity of increments property.
With these two properties and reflection symmetry, $\zeta_t$ can only be proportional to a 1D standard Brownian motion i.e., $\zeta_t=\sqrt{\kappa}B_t$, so that $\langle\zeta_t\rangle=0$ and $\langle(\zeta_t-\zeta_{t'})^2\rangle=\kappa|t-t'|$. The diffusivity $\kappa$ is the only parameter whose different values correspond to different universality classes of critical behavior. For example, $\kappa=6$ corresponds to the percolation universality class in which cluster boundaries in the continuum limit are described by SLE$_6$. In fact SLE$_\kappa$ is in general conformally \textit{covariant} under domain changes, and only for the special case of percolation with $\kappa=6$ is  conformally \textit{invariant}.

\begin{figure}[h]
\centerline{
\includegraphics[width=0.5\textwidth]{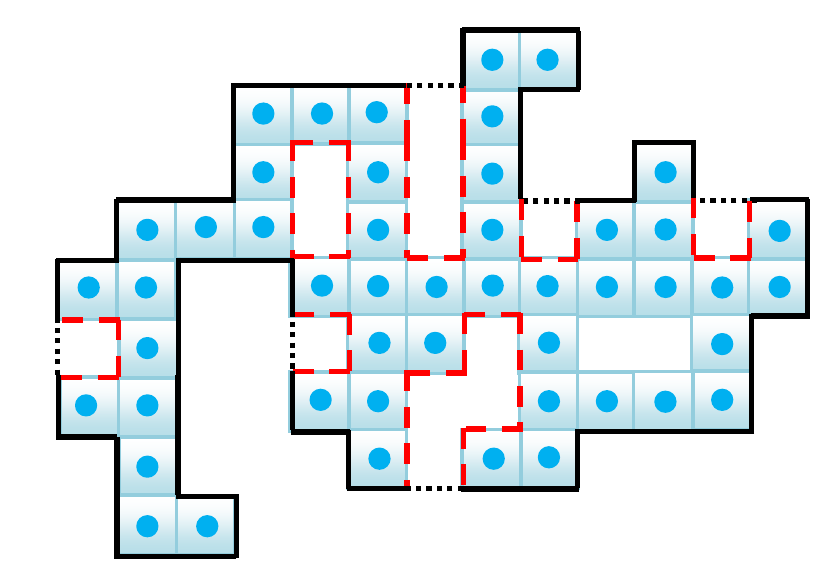}
}
 \caption{For a putative percolation cluster, the perimeter (or hull) is the union of the solid and dashed boundary lines which contains many fjords, and the external perimeter is the union of the solid and dotted boundary lines with fractal dimensions $d_{f}^{P}$ and $d_{f}^{EP}$, respectively. The external perimeter is obtained by closing off all narrow passageways. The duality relation states that $(d_{f}^{EP}-1)(d_{f}^{P}-1)=1/4$.}
\label{fig11}
\end{figure}

For $\kappa=0$, SLE curve is a vertical straight line, and when $\kappa$ 
increases the curve become more rough with fractal dimension $d_f=1+\kappa/8$ 
for $\kappa\le8$ and $2$ for $\kappa\ge8$ \cite{Beffara2008df}. For 
$0\le\kappa\le4$, the SLE curve does not intersect itself or the real axis 
(\textit{dilute phase}), while for $\kappa>4$, it intersects but does not cross 
itself and the real axis on all length scales (\textit{dense phase}). SLE is a 
space-filling curve when $\kappa\ge8$. In the dense phase with $\kappa>4$, 
there is a duality conjecture stating that the exterior frontier of an 
SLE$_\kappa$ hull looks locally as SLE$_{\tilde{\kappa}}$ with 
$\tilde{\kappa}=16/\kappa$ \cite{Duplantier2000PRL, Beffara2004SLE6}. For 
example, the external perimeter of a percolation cluster in the continuum limit 
is believed to be described by SLE$_{8/3}$ which is also expected to describe 
the scaling limit of planar self-avoiding random walk (SAW), although there is 
no complete mathematical proof yet \cite{Lawler2004PSPM}---see also 
Fig.~\ref{fig11}. 

In addition, SLE$_\kappa$ has two more special properties for $\kappa=6$ (locality) and $\kappa=8/3$ (restriction). Let $D$ be a simply connected region in $\mathbb{H}$ connected to the real axis which is at some finite distance from the origin. Consider two SLE$_\kappa$ processes from the origin to infinity, one in the domain $\mathbb{H}$ and another in the domain $\mathbb{H}\backslash D$. If these two processes have the same distribution up to the hitting time of the set $D$, then the SLE$_\kappa$ has
the locality property. Such a property is expected for the percolation cluster boundaries with $\kappa=6$ \cite{Lawler2001AM, Lawler2003JAMS}, and for no other values of $\kappa$.  Moreover, suppose that SLE$_\kappa$ with $\kappa\le 4$ has the restriction property. Then the distribution of all paths that are restricted not to hit $D$, and which are generated by SLE$_\kappa$ in $\mathbb{H}$, is the same as the distribution of all paths generated by SLE$_\kappa$ in the domain $\mathbb{H}\backslash D$. SLE$_\kappa$ has the restriction property only for  $\kappa=8/3$ and no other values of $\kappa$ \cite{Lawler2003JAMS}. We also expect such a property to hold for the continuum limit of SAWs, assuming it exists.

The winding angle between the two endpoints of a finite 2D-SAW and indeed, a broader class of critical interfaces was first studied by Duplantier and Saleur \cite{Duplantier1988PRL}. Using conformal invariance and nonrigorous Coulomb gas methods, they found that the distribution of winding angle approaches a Gaussian and they explicitly computed the winding variance as $\sim(8/g)\ln L$, where $L$ is the distance between the end points of the walk, and $g$ is a model dependent parameter which is $3/2$ for SAW. The winding angle at a single endpoint relative to the global average direction of the curve is a Gaussian with variance $\sim(4/g)\ln L$ \cite{Duplantier1988PRL}. It has been also found numerically \cite{Wieland2003Winding} that the variance in the winding at typical random points along the curve was only $1/4$ as large as the variance in the winding at the end points $\sim(1/g)\ln L$. For SLE$_\kappa$ curves, the variance in the winding angle at the end point of the curve is shown to be $\kappa\ln L$ \cite{Schramm2001SLE} (the relation between the Coulomb gas parameter $g$ and $\kappa$ can thus be given by $\kappa=4/g$).

\subsection{Scaling limit and conformal invariance of percolation}

The existence of the conformally invariant scaling limit of percolation was first conjectured in \cite{Langlands1994AMS}, based on experimental observations. This was then supported by some mathematical evidence provided for a different but related model, Voronoi percolation, which was proven~\cite{Benjamini1998Voronoi} to be invariant with respect to a conformal change of metric.
Using nonrigorous methods of CFT, Cardy could derive an exact limiting formula (\ref{CrossingProb}) for the crossing probability in a unit disc. Carleson has made an essential observation that this formula takes a particularly simple form when the domain is an equilateral triangle. In particular, for a percolation defined in an equilateral triangle $\triangle$ of side length $1$ and vertices $z_1$, $z_2$ and $z_3$, and if $z_4$ is on $(z_3,z_1)$ at distance $x\in(0,1)$ from $z_3$, then the crossing probability is simply $P_s\big((z_1,z_2),(z_3,z_4);\triangle\big)= x$. However, for years mathematicians were unable to rigorously justify the Cardy's formula.

In 2001, Smirnov \cite{Smirnov2001Proof} proved that for site percolation on the triangular lattice, the limiting crossing probability exists which is conformally invariant and satisfies Cardy’s formula. Conformal invariance of the limit means that if the domain $\Omega$, on which the percolation is defined, to be conformally mapped onto any other domain $\Omega'$, such that $z_1$ is mapped to $z'_1$, $z_2$ to $z'_2$, $z_3$ to $z'_3$ and $z_4$ to $z'_4$, then $P_s\big((z_1,z_2),(z_3,z_4);\Omega\big)=P_s\big((z'_1,z'_2),(z'_3,z'_4);\Omega'\big)$. The proof is based on the discovery of \textit{discrete harmonic} functions which encode the crossing probability and converge to conformal invariants of the domain, in the scaling limit.

Although we have been able to define the notion of a limiting crossing probability (though it may not exist at all), it was not \textit{a priori} clear until about 2000 \cite{Aizenman1997, Aizenman1998Springer, Aizenman19999DM} how to define a limiting percolation configuration i.e., a construction which does not involve limits of discrete systems. In fact, such a construction has been proposed by SLE$_\kappa$ curves as universal candidates for the scaling limits of macroscopic interfaces in 2D critical models.
\begin{figure}[h]
\centerline{
\includegraphics[width=1.0\textwidth]{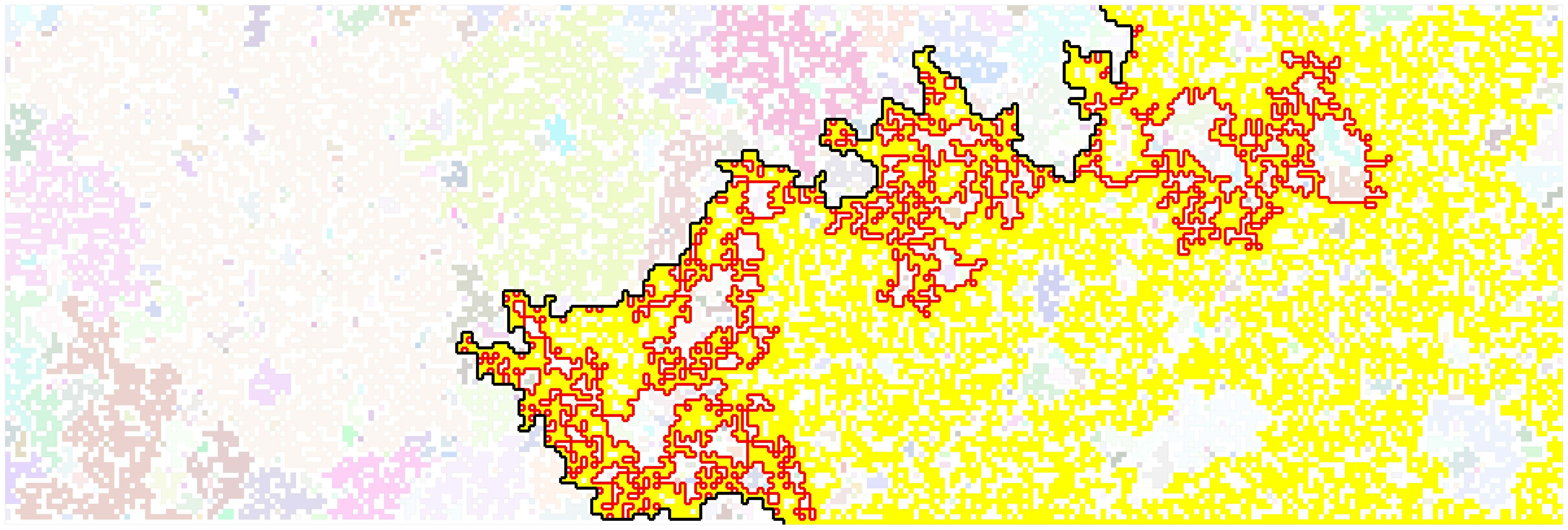}
}
 \caption{Site percolation model on a square lattice in $\mathbb{H}$. The fixed boundary condition is enforced at the lower boundary such that on the negative real half-line all the sites are unoccupied, while on the other half-line the sites are occupied. This imposes an interface (the perimeter in red) at the boundary of the spanning cluster (yellow colored) starting from the origin and ending at the upper boundary (the interface is defined uniquely by using the turn-right tie-breaking rule \cite{Saberi2009JSTAT}). The external perimeter is shown by the dark solid line. In the scaling limit, as the lattice constant goes to $0$, the perimeter and external perimeter then converge in distribution to SLE$_6$ and SLE$_{8/3}$, respectively.}
\label{fig12}
\end{figure}
In order to show the existence and conformal invariance of the scaling limit 
for percolation, one can employ either its locality property or Cardy’s formula 
for crossing probabilities to show that $\kappa=6$. Based on this observation 
Schramm conjectured \cite{Schramm2001SLE} that if percolation interface has a 
conformally invariant scaling limit, it must converge to SLE$_6$---see also 
Fig.~\ref{fig12}. Smirnov has then outlined a proof for the conformal 
invariance of the full percolation configuration. For more technical details of 
proof we refer to the original paper by Smirnov \cite{Smirnov2001Proof} or 
\cite{Riordan2006Book, Sun2011Proof, Beffara2008Is}.

As an application of this convergence result, one can prove that the critical exponents for 2D percolation  exist, and their exact values can be computed, except for $\alpha$ and $\tau$, which are nevertheless listed here for completeness:

\begin{equation}\nonumber
\alpha=-\frac{2}{3},\quad \beta=\frac{5}{36},\quad \gamma=\frac{43}{18},\quad
\tau=\frac{187}{91},\quad
 \delta=\frac{91}{5}, \quad \Delta=\frac{91}{36}, \quad \eta=\dfrac{5}{24}, \quad \nu=\dfrac{4}{3}. 
\end{equation}

\subsection{Percolation and magnetic models}

The analogy between bond percolation and conventional critical behavior in spin systems such as $q-$state Potts model, can be developed through a well-known mapping first discovered by Fortuin and Kasteleyn (FK) \cite{Fortuin1969, Fortuin1972, FK11972, FK21972}. The $q-$state Potts model is a generalization of the Ising model in which the spins at each site of a lattice can assume $q$ possible spin values. It is defined by the lattice Hamiltonian $H=-J\sum_{\langle r,r'\rangle} \delta_{s(r),s(r')}$, where the sum is over nearest neighbors and $\delta_{i,j}$ is the Kronecker delta. In the ferromagnetic case $J>0$ (we set $J=1$), the state in which all sites have the same spins minimize the energy and the system exhibits spontaneous magnetization at sufficiently low temperatures. There exists a critical temperature $T_c$ at which the system undergoes a phase transition to the disordered phase. In two dimensions for $q\le 4$ the phase transition at $T_c$ is continuous. The partition function as a function of the inverse temperature $\beta$ is $\mathcal{Z}(\beta)=\Tr \exp\big( \beta\sum_{\langle r,r'\rangle}\delta_{s(r),s(r')}\big)$, which apart from an overall unimportant constant may be rewritten as $\mathcal{Z}=\Tr\prod_{r,r'}\big((1-p)+p\delta_{s(r),s(r')}\big)$ with $p=1-e^{-\beta}$.
Now every term in this expression is associated with a bond configuration in which there exists a bond for each term $\propto p$ and there is no bond for each term $\propto (1-p)$. Sites connected by occupied bonds form clusters, and the Kronecker deltas force all the spins in each cluster to be in the same state. When the trace is taken over the spins, each cluster will have only one free spin and thus will give a factor $q$. Therefore one can write the partition function as a sum over configurations $\mathcal{C}$ of occupied bonds \cite{BaxterWu1976JPA} \begin{equation}\label{partitionfunc}
\mathcal{Z}=\sum_\mathcal{C} p^{N_b}(1-p)^{N-N_b} q^{N_\mathcal{C}}=\langle q^{N_\mathcal{C}}\rangle_{\text{percolation}},
\end{equation}
 where $N$ is the total number of bonds in the lattice, $N_b$ is the number of occupied bonds and $N_\mathcal{C}$ is the number of distinct clusters in $\mathcal{C}$. This is the random cluster (or FK) representation of the $q-$state Potts model. Clearly the limit $q\rightarrow 1$ reproduces percolation with nontrivial correlations.

 From such a correspondence (\ref{partitionfunc}), one can build a dictionary which relates the thermodynamic quantities to the geometrical properties. For example, for the Ising model with $q=2$, the spin-spin correlator is equal to the pair connectedness function of FK clusters i.e., the probability that two sites of a distance $r$ belong to the same FK cluster. The linear dimension of FK clusters diverges as the Ising correlation length, and the average (FK) cluster size diverges as the Ising susceptibility. Moreover, the average number of occupied bonds is proportional to the internal energy, and its fluctuations diverge as the Ising specific heat \cite{Hu1984PRB}. The presence of a spontaneous magnetization at $T<T_c$ reflects the appearance of an infinite cluster at $p>p_c$.

 \begin{figure}[h]
 \centerline{
 \includegraphics[width=0.6\textwidth]{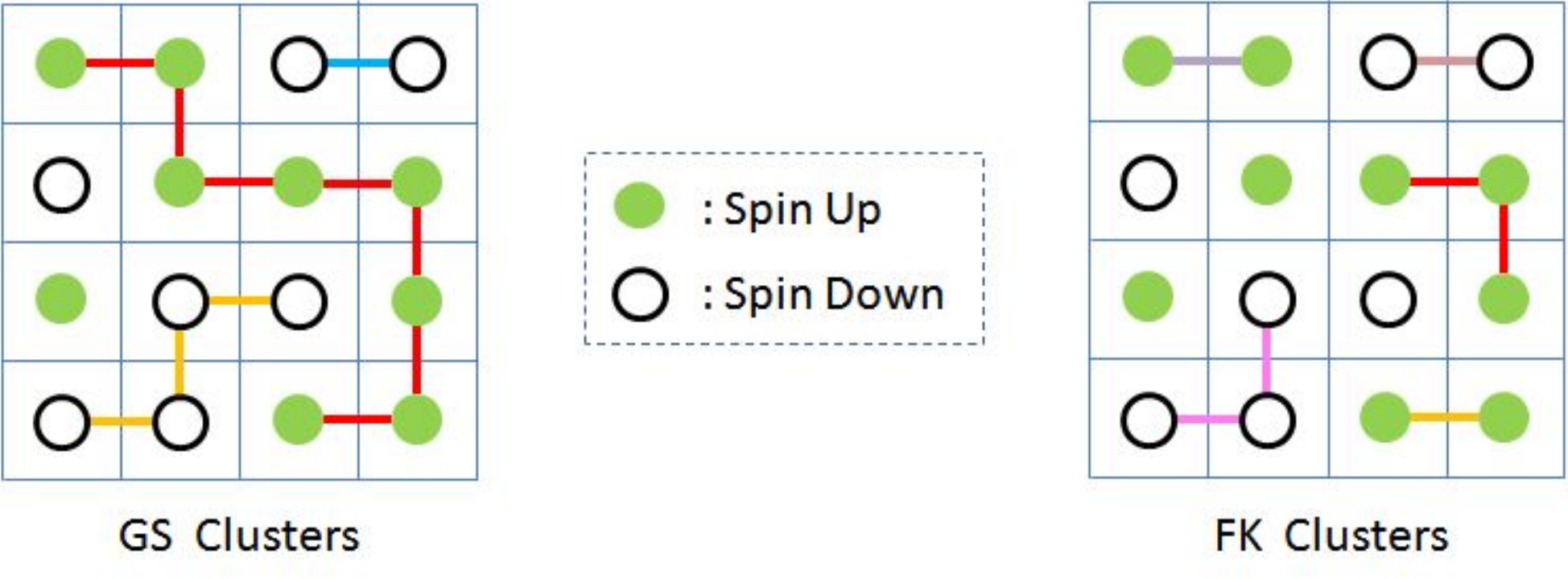}
 }
  \caption{Illustration of GS clusters (left) versus FK clusters (right) for an Ising spin configuration. Different distinct clusters are shown in different colors. A GS cluster in a spin configuration, is a set of nearest-neighbor sites of like states. An FK cluster can then be constructed from a GS cluster by randomly assigning a bond between each pair of spins with a temperature dependent probability $p:=1-e^{-2\beta}$. The critical point on the square lattice is at $\beta_c=\frac{1}{2}\ln(1+\sqrt{2})$ and $p_c=\sqrt{2}/(1+\sqrt{2})$. }
 \label{fig13}
 \end{figure}

For an Ising model on a 2D lattice with spin variables $\sigma_i=\pm 1$, an 
alternative description of the partition function can be given by 
$\mathcal{Z}=\Tr \exp(-\beta\mathcal{S})$ with the action 
$\mathcal{S}=\sum_{\langle ij\rangle}(1-\sigma_i.\sigma_j)$ which only receives 
contributions from links across which the neighboring spins are anti-aligned. 
This means that if we have two adjacent clusters of opposite spins, the 
contribution to the action is proportional to the length of the boundary of 
geometric spin (GS) clusters---see Fig.~\ref{fig13}. Thus the action can be 
considered as a sum over the self-intersecting connected admissible curves 
$\{\gamma\}$ \cite{Distler1992NPB} on the dual lattice weighted by their 
lengths $L[\gamma]$ \cite{Feynman1972Book}, i.e.,
\begin{equation}\label{Zvscurves}
\mathcal{Z}=\exp\bigg(\sum_{\{ \gamma\}}(-1)^{n(\gamma)} e^{-2\beta L[\gamma]}\bigg),
\end{equation}
where $n(\gamma)$ counts the number of intersections of the immersed curve $\gamma$. The topological term $(-1)^{n(\gamma)}$ in (\ref{Zvscurves}) is essential since it actually distinguishes between the different intrinsic topologies corresponding to a given extrinsic geometry and introduces cancellations between them to avoid over-counting of configurations. From here it is not so difficult to see that the continuum limit of this theory is a theory of free Majorana fermions that at the critical temperature $T_c$, becomes a conformal field theory with central charge $c=1/2$ \footnote{The central charge $c$ is the CFT parameter which is related to the SLE parameter $\kappa$ through $c(\kappa)=\frac{(8-3\kappa)(\kappa-6)}{2\kappa}$ \cite{Bernard2002PLB}. Central charge is invariant under the duality $\kappa\rightarrow16/\kappa$.}. More recently the existence of conformal invariance and scaling limit of the 2D Ising model was rigorously proved. It has been shown that the geometric spin (GS) cluster interfaces as well as FK cluster interfaces in a 2D Ising model strongly converge to the SLE$_3$ and SLE$_{16/3}$, respectively, in the scaling limit \cite{Smirnov2014Ising, Smirnov2010AnnMath, Smirnov2006Proc}.

\subsection{Dimensional reduction in criticality of a 3D Ising model}

The 3D Ising model has, so far, resisted an exact solution and may not be even computationally tractable \cite{Istrail2000NP}. Nevertheless much is known about its critical behavior, both analytically and numerically. \\Inspired by the formulation of the 2D Ising model as a theory of immersed curves, one may think of a possible extension of the same idea to the 3D model. There has been a lot of effort in the past \cite{Distler1992NPB, Dotsenko1987NPB3D, Sedrakyan1993PLB1, Sedrakyan1993PLB2, Fradkin1980PRD, Polyakov1981PLB, Polyakov1987HAP, Casher1985NPB, Itzykson1982NPB, Sedrakyan1984PLB, Kavalov1986PLB, Kavalov1987NPB} to reformulate the 3D Ising model in order to recast it as a string theory, i.e., as a theory of fluctuating membranes immersed in three dimensions. These attempts have been however stymied due to the difficulty in taking the continuum limit of formal sums over lattice surfaces. Part of the difficulty is that in three dimensions the topological term $(-1)^{l(\Sigma)}$(${l(\Sigma)}$ is the number of links where the closed lattice surface $\Sigma$ intersects itself) oscillates very rapidly on the length scale of the lattice spacing.

In this subsection we present a rather phenomenological approach based on application of percolation theory to give some evidence that the critical properties of a 3D Ising model are encoded in certain observables in a 2D cross section of the model. This suggests that is possible to employ the well-developed theory of immersed curves to study the Ising model in three dimensions. As discussed in the last subsection, the FK clusters in $q-$state Potts model always percolate right at the critical temperature $T_c$. The GS clusters in turn, do percolate at the same temperature $T_c$ only in two dimensions. For example, in a 3D Ising model the FK clusters percolate exactly at the Curie point $T_c$, while the percolation transition of the GS clusters, in contrast, occurs at some temperature $T_p$ well below $T_c$ \cite{Krumbhaar1974PLA} and thus the 3D GS clusters do not capture the critical properties of the model. The results of extensive Monte Carlo study of 3D Ising model on a cubic lattice shows \cite{SaberiEPL} that if one looks at an arbitrary 2D cross-section of the model, the GS clusters exhibit a percolation transition at a threshold which coincides exactly with the Curie point. It is also found numerically that the perimeter and the external perimeter of a GS cluster in a 2D cross-section at the Curie point satisfy the duality relation and their fractal dimensions and winding angle statistics are compatible, in the scaling limit, with SLE$_\kappa$ with $\kappa=5$ and $16/5$ respectively. This latter is in the same universality class as interfaces in tricritical Ising model in two dimensions. This numerical evidence may however pave the way to build a theoretical framework to understand the critical properties of the 3D Ising model.

This observation on dimensional reduction of the criticality becomes more interesting if it would be a general feature of magnetic models independent of the dimensionality and the type of microscopic interactions. In fact some recent primary results \cite{Saberi2014recent} show that even for an Ising model in four dimensions, the GS clusters in a 2D cross-section of the model percolate exactly at the critical point of the original 4D model.

\section{Percolation description of landscapes} 
\label{chap5}

Percolation theory has been extensively applied to describe the properties of both artificial and natural landscapes. Percolation properties of the correlated surfaces as a model of wide range of artificial landscapes have been discussed in subsection~\ref{corrPerc}. Our main focus in this section is mostly devoted to the statistical properties of natural landscapes and their modeling. 

The power spectrum $S$ of linear transects of Earth's topography\footnote{The power spectrum $S(k)$ is defined as the square of the coefficients in a Fourier series representation of the transect, which measures the average variation of the function at different wavelengths. For totally uncorrelated adjacent data points $S(k)$ is a constant, while for strongly correlated ones relative to points far apart, it will be large at small $k$ (long wavelengths) and small at large $k$ (short wavelengths).} have a remarkable characteristic scaling relation with the wave number $k$ as $S(k)\sim k^{-\beta_c}$ with the exponent $\beta_c=2$, over a wide range of scales \cite{Vening1951PKNAWS, Mandelbrot1975PNAS, Sayles1978Nature, Newman1990GJI}. Similar scaling relations have been identified in Earth’s bathymetry (i.e., the underwater equivalent to topography) \cite{Bell1975DSR}, the topography of natural rock surfaces \cite{Brown1985JGR}, and the topography of Venus \cite{Kucinskas1992JGR}. Such a power-law spectrum in the topography is responsible for the appearance of various self-similar patterns on Earth, e.g., fractal coastlines \cite{Mandelbrot1967Sci}, the radiation fields of volcanoes \cite{Harvey2002Fractals, Lovejoy2003IJRS}, crustal density and gravity \cite{Pilkington2004GRL}, geomagnetism \cite{Pecknold2001GJI}, and surface hydrology such as in the river basin geomorphology \cite{Rodriguez1997Book}. Although environmental parameters such as erosion seem to play an important role in shaping the coastlines, drainage basins and watersheds, the observation of scale-invariant topography on Venus, however, indicates
that fractal topography can be formed without erosion.
\\The exponent $\beta_c$ is related to the Hurst exponent $H$ in fractional Brownian motion (fBm) via $\beta_c= 2H+1$, thus suggesting $H\simeq0.5$ for Earth's topography. However, further surveys based on the fBm model \cite{Mandelbrot1975PNAS} of topography or bathymetry revealed a more complex multifractal structure of Earth’s morphology giving rise to distinct scaling properties of oceans, continents, and continental margins describes by $H=0.46, 0.66$ and $0.77$, respectively \cite{Gagnon2006NPG}.

The other characteristic feature of Earth's topography is its bimodal distribution \cite{Wegener1929Book} which reflects the topographic dichotomy of continents and ocean basins. It has a clear discrepancy with Gaussian models of topography, also consistent with the Mandelbrot's observation \cite{Mandelbrot1983Book}.  The positive correlation between elevation and slope seen on Earth (i.e., the steepness increases with the height) is not also predicted by a model with a Gaussian distribution, implying that the global topography of Earth is not easily amenable to modeling.

Nevertheless, percolation theory has been recently applied to describe the global topography of Earth \cite{Saberi2013PRL}, in which the critical point indicates the present mean sea level. Moreover, different models based on percolation theory have been proposed to describe the statistical properties of regional features on Earth such as coastlines  \cite{Sapoval2003, Morais2011PRECoast}, river basins and drainage networks \cite{Maritan1996Science, Maritan1997PRL, Maritan1998JSP, Maritan1997PRE, Hergarten2001PRL, Stark1991Nature}, and watersheds \cite{Fehr2009Watershed, Fehr2012WatershedPRE, Fehr2011PRLWatersheds, Herrmann2011PP}. 
Percolation theory has been also successful to help understand other phenomena on Earth. It has been demonstrated \cite{Golden1998Ice} that sea ice exhibits a percolation transition at a critical temperature above which brine carrying heat and nutrients can move through the ice, whereas for colder temperatures the ice is impermeable. Percolation also serves as an attractive mechanism to explain core formation in Earth \cite{Shannon1998Science, Mann2008Core}.

\subsection{Fractal geometry of coastlines}

Coastlines were among the first natural systems that have been quantitatively characterized when Mandelbrot computationally analyzed their fractal geometry \cite{Mandelbrot1967Sci}. In fact the geometrical irregularity of the coastlines helps damping the sea waves and decreasing the average wave amplitude. Affected by the sea eroding power, an irregular morphology evolves at the rocky coast until a self-stabilization with the wave amplitude is established. A simple model of such stabilization has been studied \cite{Sapoval2003} in which the fractal geometry of the coastline plays the role of a morphological attractor directly related to percolation geometry. Dynamics of the model spontaneously leads to a stationary fractal geometry with a dimension very close to $4/3$ independent of the initial morphology, in agreement with that is observed on real coasts \cite{Mandelbrot1967Sci, Sapoval1989fractals}. This fractal dimension is also consistent with that of the external perimeter of the spanning cluster in a 2D critical percolation. Two general erosion mechanisms are considered in the 2D model, i.e., a rapid mechanical erosion and a slow chemical weakening. It has been shown that when the model involves both processes, a dynamic equilibrium is reached that changes the shape of the coast but preserves its fractal properties.

The effect of spatial long-range correlations in the lithology of coastal landscapes on the fractal properties of the coastlines has then been addressed in \cite{Morais2011PRECoast}. In fact, due to the \textit{endogenic} processes like volcanic activity,
earthquakes, and tectonic processes originating within Earth that are mainly responsible for the very long-wavelength topography of Earth’s surface, one naturally expects  that lithological properties of coastal landscapes would be in general heterogeneous as well as long-range correlated in space. Moreover, a multitude of fractal dimensions
has been measured for real coastlines of different landscapes \cite{Richardson1961yearbook}. Thus self-similar geometry of coastlines should emerge from an intricate interplay between these landscape properties and the sea force. The results of a simple invasion model \cite{Morais2011PRECoast} indicates that a critical sea force $f_c$ exists at which the coastline exhibits self-similarity with fractal dimension depending on Hurst exponent. The dominant $4/3$ fractal dimension was obtained for uncorrelated landscapes. For $f<f_c$ the coastline is rough but not fractal and the eroding process stops after some time, while for $f>f_c$, erosion is perpetual leading to a self-affine coastline which  belongs to the Kardar-Parisi-Zhang (KPZ) \cite{Kardar1986KPZ}
universality class.

As discussed in Sec.~\ref{chap4}, the external perimeter of critical  percolation clusters with fractal dimension $4/3$ are proven to have a conformally invariant limit described by SLE$_{8/3}$. Some numerical evidence of such a strong symmetry has been also reported for rocky coastlines with fractal dimension $4/3$ \cite{Boffetta2008GRL}. 
These coastlines are therefore shown to be statistically equivalent to the external perimeter of percolation clusters or that of planar random walk.
The conformal invariance can then be used to predict the statistics of the flux of pollutants diffusing over shorelines. This flux has been characterized by a strongly intermittent spatial distribution which can vary dramatically between locations just a few hundred meters apart.\\Strong evidence of conformal invariance property has been also presented for the iso-height lines (like coastlines) of both artificial landscapes in the KPZ universality class \cite{Saberi2008PRE1, Saberi2009PRE0, Saberi2010PRE0} and experimentally grown surfaces \cite{Saberi2008PRL}. In the KPZ universality class, the iso-height lines are characterized by a fractal dimension of $4/3$ with the same conformal invariant properties as the external perimeter of critical percolation clusters, compatible with SLE$_{8/3}$ curves in the scaling limit. Such an analogy may lead to an alternative description of the coastlines.

\subsection{Statistical properties of watersheds}

The watershed is defined as the line which separates adjacent drainage 
basins---see Fig.~\ref{fig5}. Based on observation of natural watersheds, it 
has been claimed that they should have a fractal structure 
\cite{Breyer1992Geo}. The self-similarity of watersheds has then been justified 
numerically for both natural and artificial landscapes  
\cite{Fehr2009Watershed, Fehr2011PRLWatersheds, Fehr2011PREWatershed}. 
Watersheds have been shown to be related to a family of curves appearing in 
different contexts, e.g., bridge percolation \cite{Schrenk2012Fracturing}, 
polymers in strongly disordered media \cite{Porto1997PRL}, optimal path cracks 
\cite{Andrade2009PRLFracture}, and fracturing process 
\cite{Moreira2012PRLFracture}.
To determine the watershed lines on real or artificial landscapes which are usually in the form of Digital Elevation Maps (DEM), consisting of discretized elevation fields, one can use an iterative application of an invasion percolation procedure \cite{Fehr2009Watershed}.

The fractal dimensions of watershed lines in 2D and 3D were estimated \cite{Fehr2012WatershedPRE} to be $d_f=1.2168\pm 0.0005$ and $2.487\pm0.003$, respectively, for uncorrelated artificial landscapes. In two dimensions, however, the measured fractal dimensions for natural landscapes obtained from data provided by satellite imagery \cite{Farr2007Radar}, fall into the range $1.10\le d_f\le1.15$. This may imply the necessity of considering spatial correlations in computations. When the long-range correlations  characterized by the Hurst exponent $H$ were introduced \cite{Fehr2011PREWatershed}, a monotonic decrease of $d_f$ with $H$ has been observed, and the agreement with the observation achieved for $0.3<H<0.5$ (although this range of $H$ seems to be out of that is observed for continents and continental margins \cite{Gagnon2006NPG}). Moreover, it has been shown \cite{Fehr2011PRLWatersheds} that small and localized perturbations like landslides or tectonic activities, can have a large and non-local impact on the shape of watersheds.
It is also discussed in \cite{Herrmann2011PP} that the fractal dimension obtained in 2D for uncorrelated artificial landscapes is intriguingly close to the fractal dimension of the largest cluster boundary in two models of explosive percolation on a lattice, i.e., the \textit{largest cluster} and \textit{Gaussian} models \cite{Hermann2010Explosive}.

Watersheds are shown \cite{Daryaei2012PRL} to be among the rare examples of physical systems described by SLE$_\kappa$ curves with $\kappa<2$. It has been numerically shown that, in the scaling limit, the watershed line exhibits conformally invariant properties compatible with SLE$_\kappa$ with $\kappa=1.734\pm0.005$.

\subsection{The present mean-sea level on Earth}

The ubiquitous scale invariant features on Earth have endowed theoretical interest on the assumption that they may reveal prevalence of some
underlying feature \cite{Bak1996book, Jensen1998Book, Sornette2000Book}.
This is still an open question if there exists a clear relationship between the quantitative properties of 
landscapes and the dominant geomorphologic processes that originate them. Although such a relationship is established for some of regional features, the global topography in comparison, has received less attention.

The appearance of scale and conformal invariance property in statistical models like percolation is a specific feature of criticality. This can be regarded as a motivating issue to search for an underlying mechanism that possibly explains the emergence of fractal geometries on various landscapes. For instance, it has been shown for an ensemble of experimentally grown surfaces \cite{SaberiAPL} that there exists a critical level height at which a percolation transition occurs. This may elucidate the earlier observation \cite{Saberi2008PRL} of conformal invariant iso-height lines on these samples.

\begin{figure}[h]
\centerline{
\includegraphics[width=0.8\textwidth]{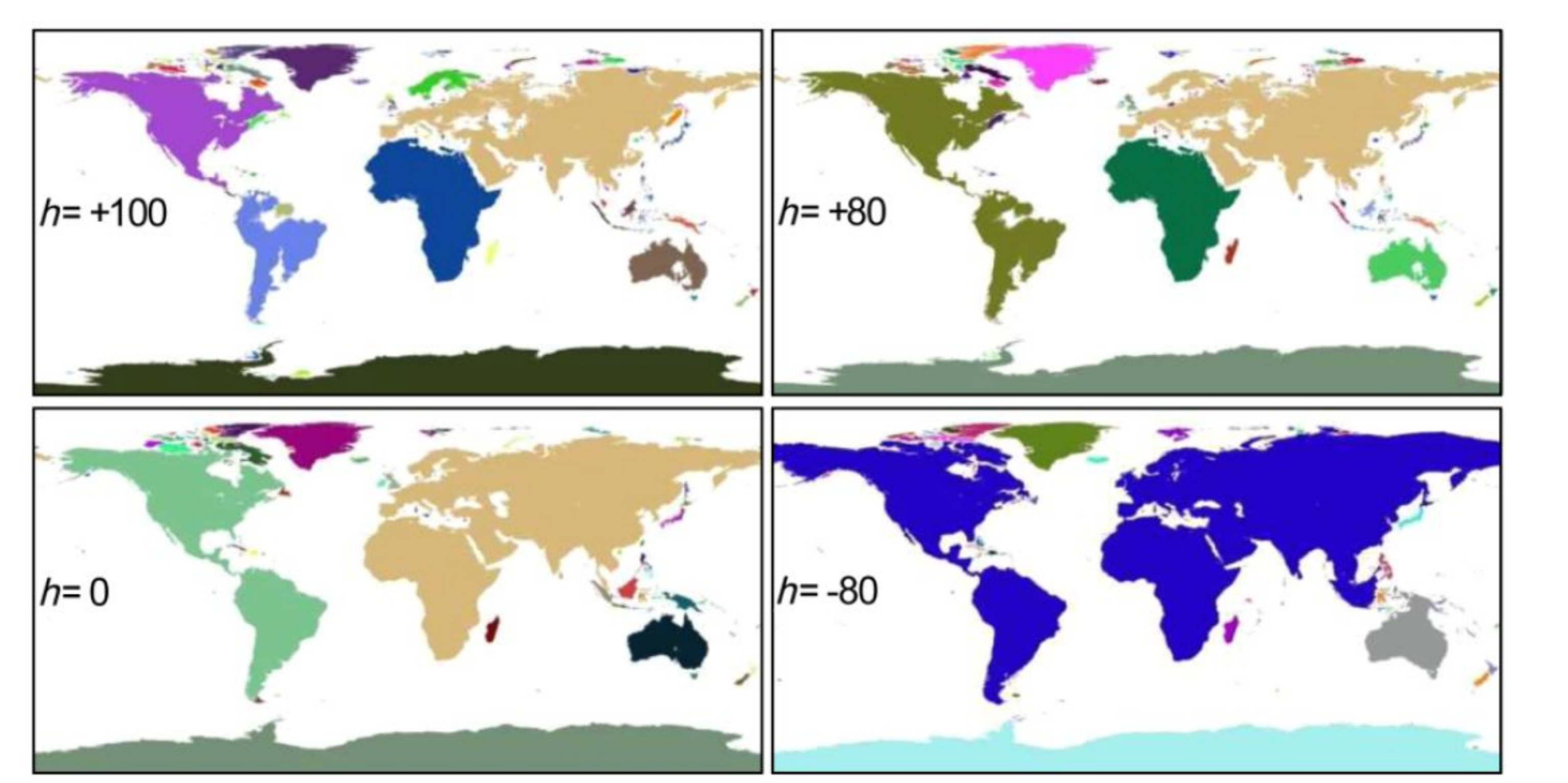}
}

 \caption{Schematic illustration of the continental aggregation by decreasing the sea level from top ($h=+100$ m) to bottom ($h=-80$ m). This shows a remarkable percolation transition at the present mean sea level around which the major parts of the landmass join together.}
\label{fig14}
\end{figure}

A percolation description of the global topography of Earth is recently presented \cite{Saberi2013PRL} in which a dynamic geoid-like level is
defined as an equipotential spherical surface as a counterpart of the 
percolation parameter. When the hypothetical water level is decreased from the 
highest to lowest available heights on Earth, there occurs a geometrical phase 
transition at a certain critical level $h_c$ around which the most parts of 
landmass join together---see Fig.~\ref{fig14}.  
The most remarkable observation is that the critical level $h_c$ coincides with the present mean sea level $h=0$ on Earth. The criticality of the current sea level justifies the appearance of the scale (and conformal) invariant features on Earth. This may also uncover the important role that is played by water on Earth and shed new light on the tectonic plate motion.

\begin{figure}[h]
\centerline{
\includegraphics[width=0.5\textwidth]{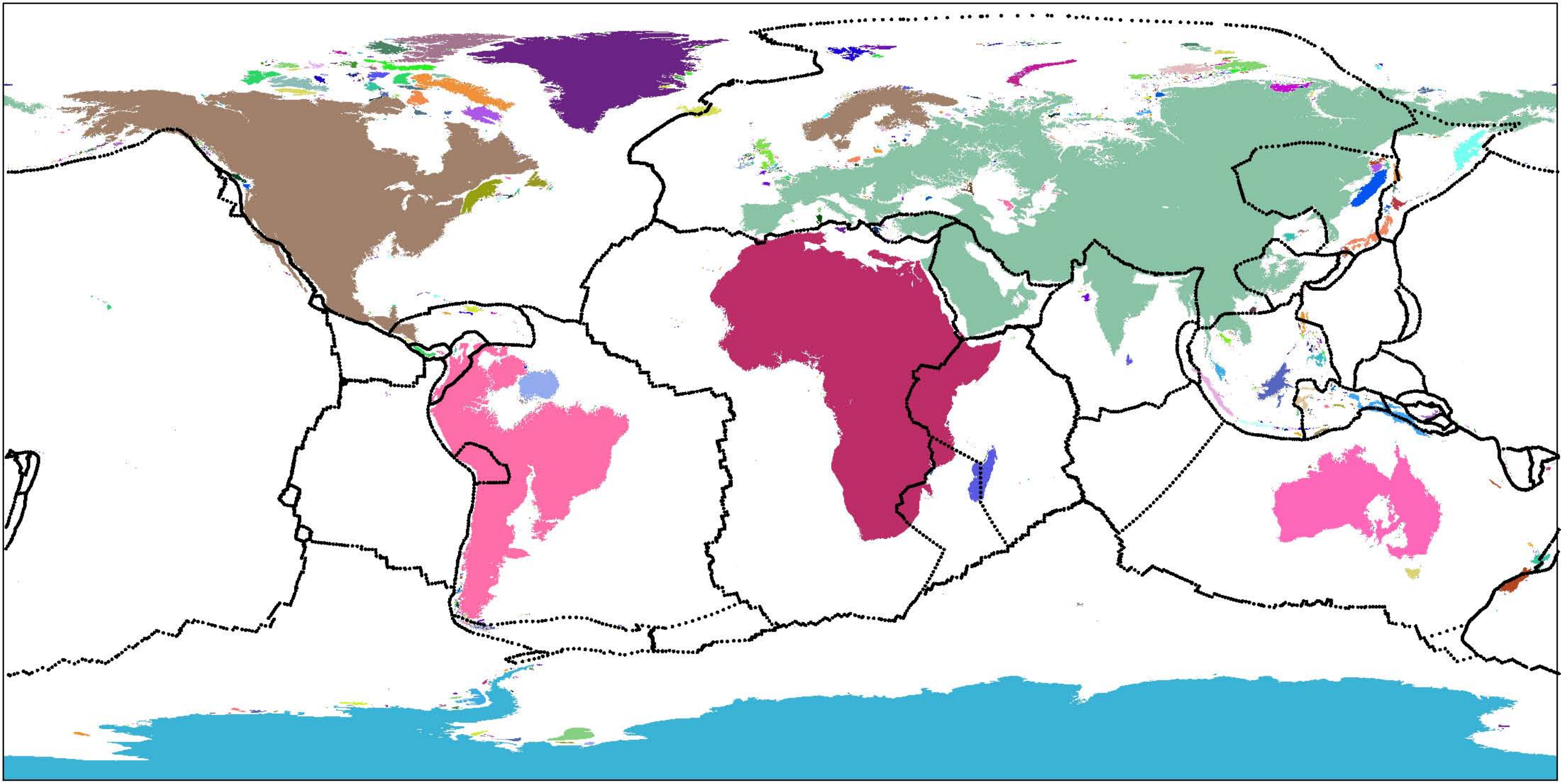}
}
 \caption{Map of tectonic plates compared with different disjoint islands for the sea level at $h=+100$~m. Every disjoint landmass is approximately surrounded by a major plate boundary. }
\label{fig15}
\end{figure}

According to the plate tectonic theory, the outer portion of Earth is made up 
of a number of distinct \textit{plates} (Fig.~\ref{fig15}) which move relative 
to each other. This motion is responsible for the major topographical features 
such as creation of oceans and pushing up mountain ranges. 
The open question motivated by this work is whether such an observed criticality plays the role of a geometric attractor for tectonic motions through geological time.

\section{Conclusions}
\label{conclusion}

Through the current review we have outlined some basic properties and recent advances of percolation theory as well as some of its recent applications which can be summed up as the following statements. Percolation theory and its applications span a wide area of science ranging from social and network sciences to string theory and particle physics as well as various branches of probability theory in mathematics. The percolation models are mostly governed by very simple rules yet with a fascinating mathematical structure and fundamental features. It has an uncorrelated microscopic structure, but a long-ranged correlated geometry can emerge which governs the system. All its properties can be described by a number of geometric observables with some characterizing features distinguishing between subcritical, critical and supercritical phases. The criticality has a rich fractal structure and remarkable underlying scaling laws defined by some universal critical exponents. Percolation theory simultaneously benefits from exact conjectures raised by physical insights on the one hand, and rigorous mathematical proofs on the other hand. It has a very robust nature against small perturbations but ready to play a role in a completely different scene under sufficiently large modifications. With no interactions, it may be discovered in the heart of strongly interacting systems \footnote{For example, in 2D turbulence, a paradigmatic example of strongly interacting
non-equilibrium system, it is numerically shown~\cite{Bernard2006Nature} that the statistics of vorticity clusters is remarkably close to that of
critical percolation.}. 
While it is applied to formulate the critical behavior of a system in terms of some appropriate observables, it can be viewed at the same time as another theory in a lower dimension describing the same system with appropriately well defined observables~\cite{SaberiEPL}. Many exact results have been obtained but there still exist many open challenges in the field \cite{Ziff2014Recent}.

\section*{Acknowledgement}

I would like to thank J. Cardy, H.J. Herrmann, J. Nagler, M. Sahimi, R. Ziff and 
especially D. Stauffer for 
critical reading of the manuscript and their very valuable comments. I also would 
like to thank H. Dashti-Naseabadi for his kind helps. I wish to express my gratitude for the hospitality of ICTP in 
Trieste, Italy, where some of the work was done. 
I also acknowledge partial financial supports by the Iran National Science Foundation (INSF), and University of Tehran.

\section*{References}

\bibliographystyle{plainnat}
\bibliography{Percolation}

\end{document}